\documentclass[twocolumn,showpacs,preprintnumbers,amsmath,amssymb,superscriptaddress]{revtex4}
\usepackage{amsmath,amsfonts,amssymb}
\usepackage[english]{babel} 
\usepackage[latin1]{inputenc} 
\usepackage[T1]{fontenc}
\usepackage{color}
\usepackage{float}
\usepackage{verbatim}
\usepackage{graphicx}
\usepackage{bm}
\usepackage{mathtools}

\def \equi#1{\mathrel{\mathop{\kern 0pt\sim}\limits_{#1}}}

\newcommand{\e}[1]{\; e^{#1} \;}
\newcommand{\erfc}[1]{\; \mathrm{erfc}\left(#1\right) \;}
\newcommand{\erf}[1]{\; \mathrm{erf}\left(#1\right) \;}

\begin{document}

\title{Mean perimeter of the convex hull of a random walk in a semi-infinite medium}

\author{Marie Chupeau}
\affiliation{Laboratoire de Physique Th\'eorique de la Mati\`ere Condens\'ee (UMR CNRS 7600), Universit\'e Pierre et Marie Curie, 4 Place Jussieu, 75255
Paris Cedex France}

\date{\today}
\author{Olivier B\'enichou}
\affiliation{Laboratoire de Physique Th\'eorique de la Mati\`ere Condens\'ee (UMR CNRS 7600), Universit\'e Pierre et Marie Curie, 4 Place Jussieu, 75255
Paris Cedex France}

\author{Satya N. Majumdar}
\affiliation{CNRS, Laboratoire de Physique Th\'eorique et Mod\`eles Statistiques, Universit\'e Paris-Sud, 91405 Orsay Cedex, France}

\begin{abstract}
We study various properties of the convex hull of a planar Brownian motion, defined as the minimum convex polygon enclosing the trajectory, in the presence of an infinite reflecting wall. Recently, in a Rapid Communication [Phys. Rev. E \textbf{91}, 050104(R) (2015)], we announced that the mean perimeter of the convex hull at time $t$, rescaled by $\sqrt{Dt}$, is a non-monotonous function of the initial distance to the wall. In the present article, we first give all the details of the derivation of this mean rescaled perimeter, in particular its value when starting from the wall and near the wall. We then determine the physical mechanism underlying this surprising non-monotonicity of the mean rescaled perimeter by analyzing the impact of the wall on two complementary parts of the convex hull. Finally, we provide a further quantification of the convex hull by determining the mean length of the portion of the reflecting wall visited by the Brownian motion as a function of the initial distance to the wall.
\end{abstract}


\maketitle

\section{Introduction}

Characterizing the territory covered by a Brownian motion in two dimensions is a natural question, both in the context of the theoretical study of a bidimensional Brownian motion and in ecology, where the trajectories of foraging animals are often satisfactorily modeled by a Brownian motion~\cite{Berg:1983,Bartumeus:2005}. Indeed, it can be necessary to estimate the home range of an animal, defined as the two dimensional space over which an animal moves around over a fixed period of time~\cite{Murphy:1992}. Ecologists frequently estimate this home range by computing the convex hull of the trajectory of the animal, i.e. the minimal convex polygon enclosing the trajectory~\cite{Worton:1995,Giuggioli:2011}, and calculating its perimeter or area.

The perimeter and the area of the convex hull of isotropic 2D stochastic processes have been extensively studied both in the physics~\cite{MajumdarPRL09,Majumdar:2010,Reymbaut:2011,Dumonteil:2013,Lukovic:2013,Randon:2013,Randon:2014} and mathematics literatures~\cite{Takacs:1980,ElBachir:1983,Letac:1993,Biane:2011,Eldan:2014,Kampf:2012,Kabluchko:2014}. Beyond the basic calculation of the mean perimeter and area in the case of a Brownian motion, the literature gathers various extensions, for example the case of $N$ independent Brownian motions~\cite{MajumdarPRL09,Majumdar:2010}, random acceleration processes~\cite{Reymbaut:2011}, branching Brownian motion with absorption~\cite{Dumonteil:2013} and anomalous diffusion processes~\cite{Lukovic:2013}. When the process is isotropic, the calculation of the mean perimeter
and area of the convex hull can be conveniently carried out by studying the extremal statistics of the corresponding
one-dimensional radial process, as presented in \cite{MajumdarPRL09,Majumdar:2010}.

\begin{figure}
\centering
\includegraphics[width=130pt]{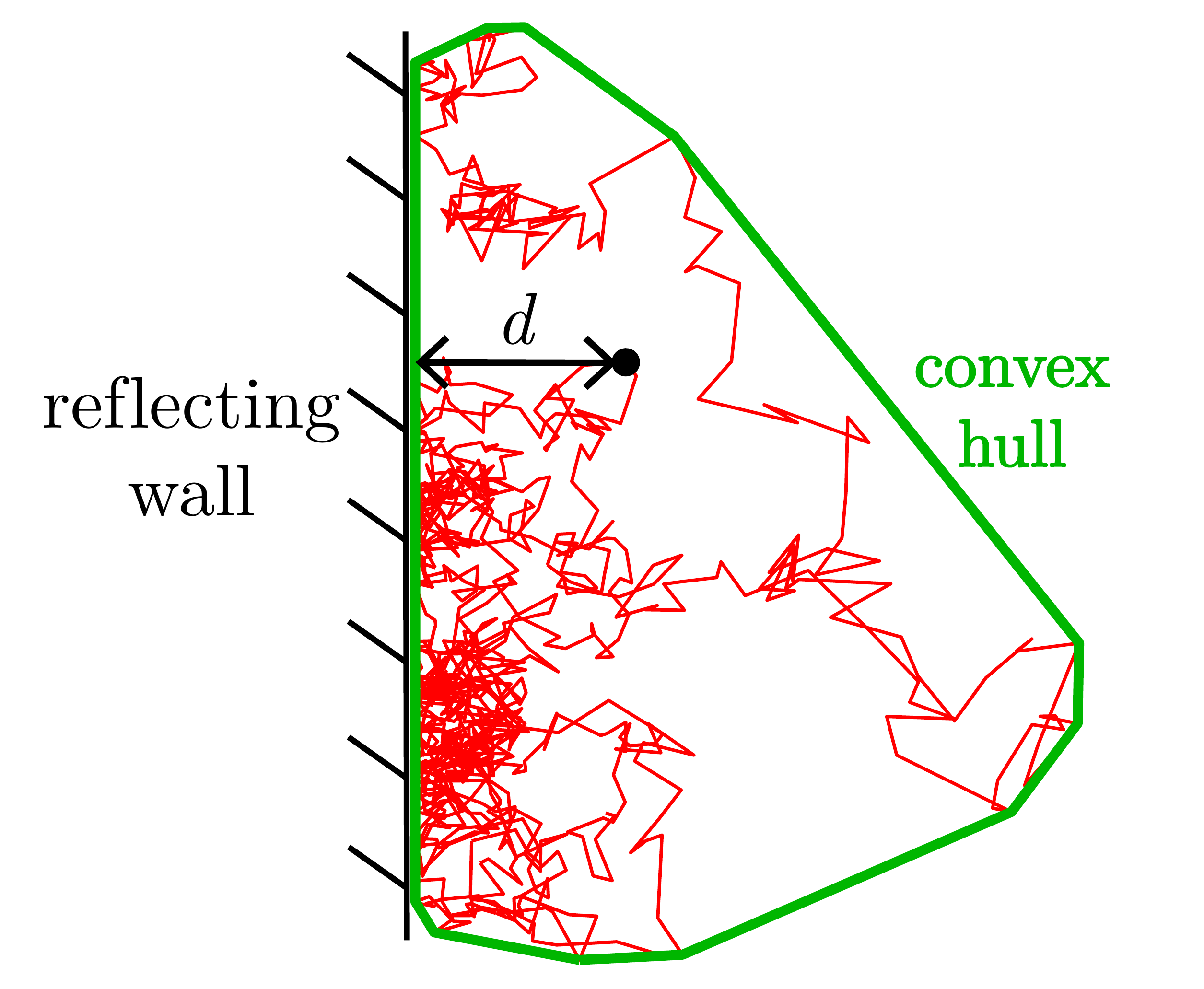}
\caption{Minimal model of a Brownian motion in the presence of a reflecting infinite wall, that starts at a distance $d$ from the wall. The trajectory is represented by the thin red path, and the convex hull of the Brownian motion, defined as the minimum convex polygon enclosing the trajectory, is the thick green polygon.}
\label{def_probleme}
\end{figure}

So far, all these studies focused on unconfined bidimensional processes. However, the natural environment of animals is hardly ever unlimited, their displacements being constrained by natural or human-built obstacles, such as littorals, mountains, urban areas, roads... Moreover, the question of the impact of a confinement on the characteristics of the convex hull of a Brownian motion is also essential in the theoretical study of Brownian motion. Recently, a Rapid Communication \cite{Chupeau:2015b} has addressed this question by considering the minimal model of a single planar Brownian motion in the presence of a reflecting infinite wall that confines the Brownian motion in a half-space (see Fig.~\ref{def_probleme}). This confinement, though simple, is suited to model a river or a road that cannot be crossed. The presence of the reflecting wall has a non-trivial effect on the mean perimeter of the convex hull. Indeed, it was shown that this confinement produces a surprising non-monotonicity of the mean perimeter of the convex hull at time $t$, rescaled by $\sqrt{Dt}$, with respect to the initial rescaled distance to the wall $x$, and a singularity for a Brownian motion starting very close to the wall (\mbox{$x\ll 1$}). 

In the present paper, (i) we give all the details of the derivation of the mean rescaled perimeter, (ii) go further than this calculation, both by qualitatively and quantitatively studying the mechanism that produces the non-monotonicity of the mean rescaled perimeter, and (iii) focus on an additional observable, the extension of the segment of the wall that has been visited by the Brownian motion.
More precisely, in Sec.~II, we determine the mean rescaled perimeter of the convex hull of a Brownian motion starting at a distance $d$ from an infinite reflecting wall, and analyze it for Brownian motions starting from the wall and near the wall. We also provide details on the non-trivial question of the numerical evaluation of the analytical expression of the mean rescaled perimeter, and on the numerical simulations. In Sec.~III, we determine the physical mechanisms underlying the non-monotonicity of the mean rescaled perimeter by studying the impact of the wall on two complementary parts of the convex hull. Finally, in Sec.~IV, we focus on a subset of the convex hull, the extension of the visited points on the wall, that represents a further quantification of the convex hull, and analyze its dependence on the initial distance to the wall.  

\section{Determination and analysis of the mean perimeter of the convex hull}

\subsection{Determination of the mean perimeter}

\begin{figure}[h]
\centering
\includegraphics[width=120pt]{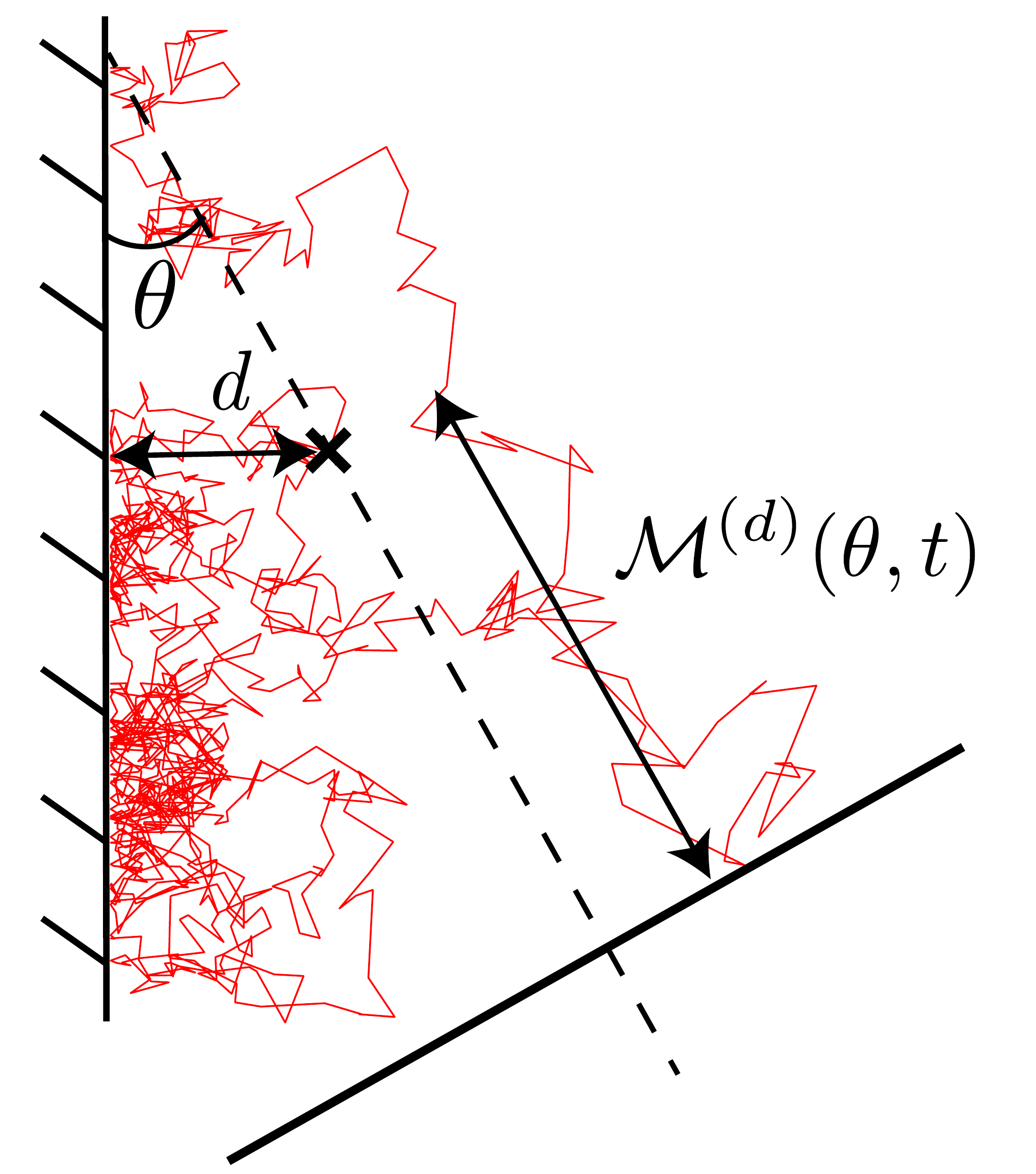}
\caption{Definition of the maximum \mbox{$\mathcal{M}^{(d)}(\theta,t)$} of the trajectory in the direction $\theta$ at time $t$ for a Brownian motion starting at a distance $d$ from the reflecting wall.}
\label{def_max}
\end{figure}

\begin{figure}[h]
\centering
\includegraphics[width=100pt]{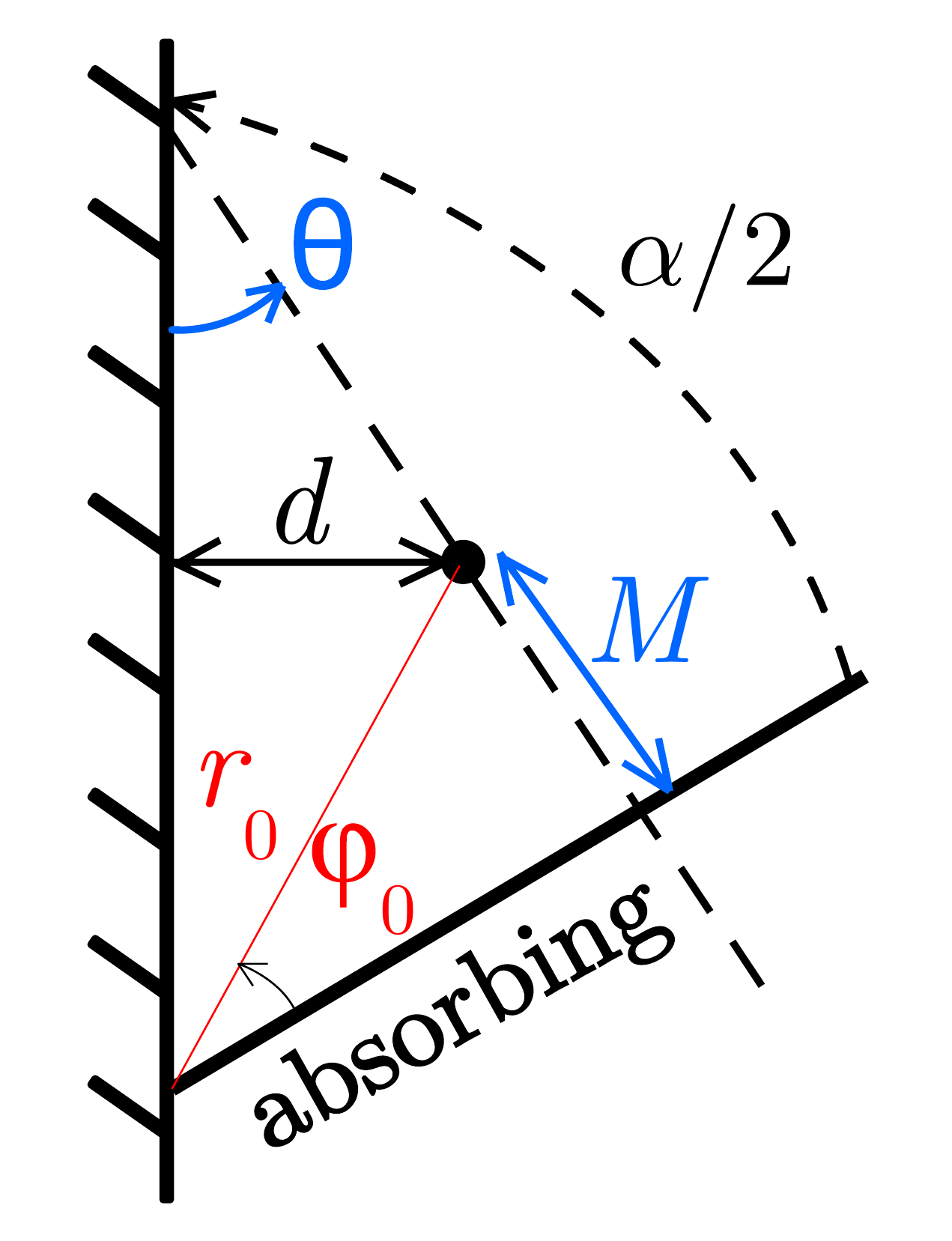}
\caption{Definition of the geometry.}
\label{geom}
\end{figure}

\begin{figure}[h]
\centering
\includegraphics[width=200pt]{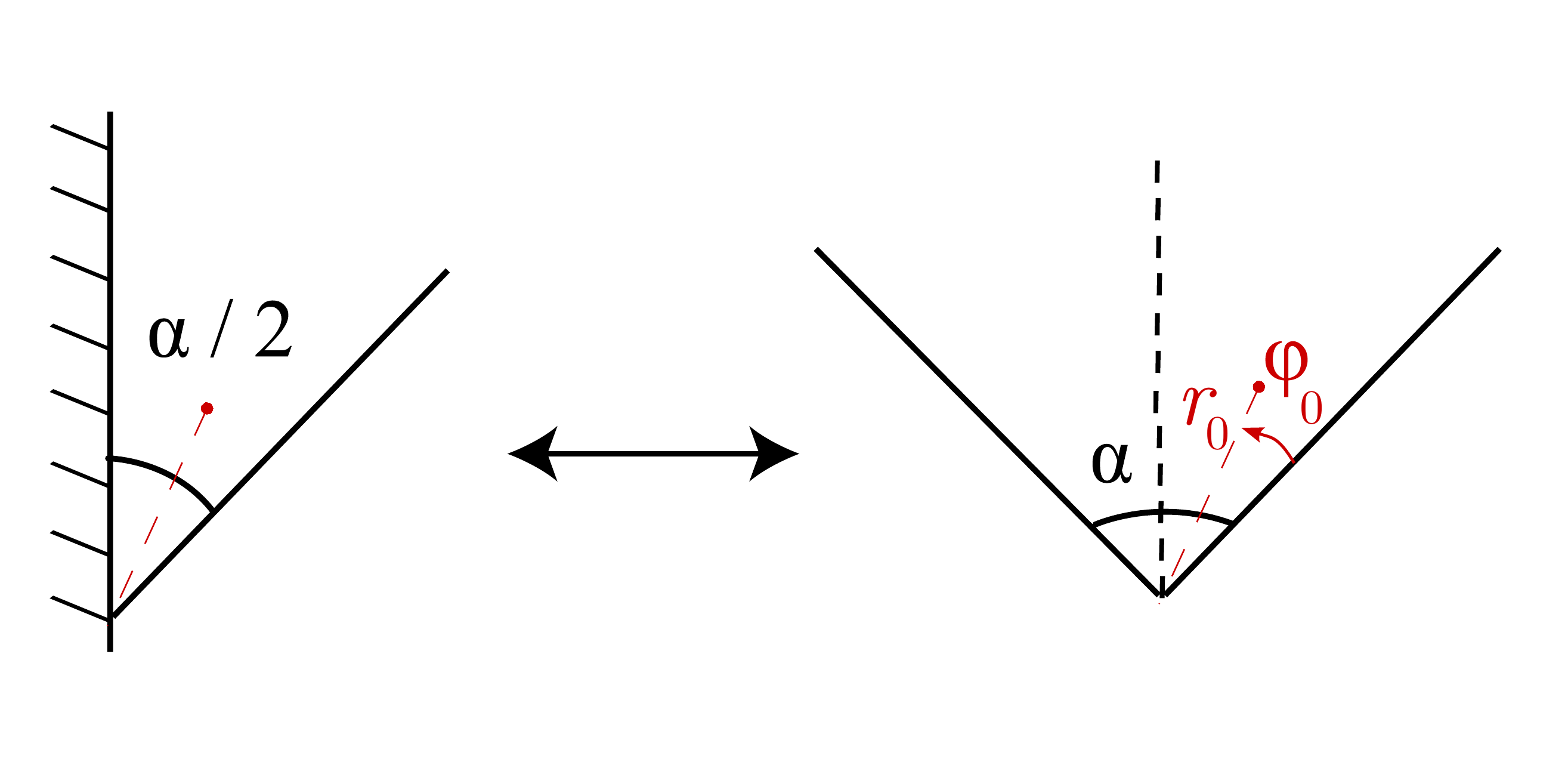}
\caption{Equivalence between a wedge with an absorbing and a reflecting edges and a wedge of double top angle with two absorbing edges. The starting point is at the polar coordinates \mbox{$(r_0,\varphi_0)$}.}
\label{wedgeequiv}
\end{figure}

The first step to calculate the mean perimeter of the convex hull \mbox{$\langle L^{(d)}(t) \rangle$} at time $t$ for a Brownian motion starting at a distance $d$ from the reflecting wall consists of relating this quantity with the mean maximum \mbox{$\langle \mathcal{M}^{(d)}(\theta,t) \rangle$} of the trajectory in the direction $\theta$ using Cauchy formula \cite{MajumdarPRL09}
\begin{equation}\label{Cauchy}
\langle L^{(d)}(t) \rangle=\int_0^{2 \pi} d\theta \langle \mathcal{M}^{(d)}(\theta,t) \rangle.
\end{equation}
The maximum of the trajectory in a direction \mbox{$\theta\in [0,2\pi]$} corresponds to the minimal distance between the starting point and the lines orthogonal to the direction $\theta$ that do not touch the trajectory (see Fig.~\ref{def_max}). We denote by \mbox{$F(\mathcal{M}^{(d)}(\theta,t)=M)$} the probability density of the extension \mbox{$\mathcal{M}^{(d)}(\theta,t)$}, such that
\begin{equation}\label{defM2}
\langle \mathcal{M}^{(d)}(\theta,t) \rangle =\int_0^{+\infty} dM M \; F(M), 
\end{equation}
and by \mbox{$S^{(d)}(t|M,\theta)$} the survival probability at time $t$ of a Brownian motion starting at a distance $d$ from the wall, in the presence of an infinite absorbing wall orthogonal to the direction $\theta$ located at distance $M$ from the starting point  (see Fig. \ref{geom}).
The probability for a Brownian motion to have a maximal extension larger than $M$ in the direction $\theta$ at time $t$ is the probability to have been absorbed before time $t$ by the absorbing wall mentioned above. The absorption probability, which is equal to \mbox{$1-S^{(d)}(t|M,\theta)$}, is
\begin{equation}
1-S^{(d)}(t|M,\theta)=\int_M^{+\infty} dm \; F(m).
\end{equation}
This leads to
\begin{equation}
F(M)=-\dfrac{d}{dM}\left(1-S^{(d)}(t|M,\theta)\right).
\end{equation}
An integration by parts, taking into account that the survival probability at time $t$ is one when the absorbing wall is far from the starting point (\mbox{$M \to +\infty$}), yields
\begin{equation}\label{defmax}
\langle \mathcal{M}^{(d)}(\theta,t) \rangle=\int_0^{+\infty} dM \left(1-S^{(d)}(t|M,\theta)\right).
\end{equation}
In this geometry, the directions \mbox{$[-\pi/2,\pi/2]$} are equivalent to the directions \mbox{$[\pi/2,3\pi/2]$}. Hence, in what follows, we will restrict our calculations to the range \mbox{$\theta\in[-\pi/2,\pi/2]$}. Using Eq.~\eqref{Cauchy}, we finally obtain 
\begin{equation}\label{defperim}
\langle L^{(d)}(t) \rangle = 2 \int_{-\pi/2}^{\pi/2} d\theta \int_0^{+\infty} dM \left( 1-S^{(d)}(t|M,\theta) \right).
\end{equation}
The calculation of \mbox{$\langle L^{(d)}(t) \rangle$} then involves the determination of the survival probability in a wedge with one absorbing edge and one reflecting edge, previously defined. This wedge is equivalent to a wedge with two absorbing edges of double top angle, as illustrated in Fig.~\ref{wedgeequiv}. The survival probability at time $t$ in such a wedge for a starting point at a distance $r_0$ of the apex and parametrized by an angle $\varphi_0$ (see Fig.~\ref{wedgeequiv}) is shown in Appendix~\ref{Survival} to be
\begin{align}\label{survie}
&S(t|r_0,\varphi_0)=\dfrac{r_0}{\sqrt{\pi Dt}} \;e^{-\frac{r_0^2}{8Dt}} \sum\limits_{m=0}^{+\infty} \dfrac{\sin\left(\frac{(2m+1)\pi \varphi_0}{\alpha}\right)}{2m+1} \nonumber \\
&\quad \times \left[ I_{\frac{(2m+1)\pi}{2 \alpha}-\frac{1}{2}} \left(\frac{r_0^2}{8Dt}\right)+ I_{\frac{(2m+1)\pi}{2 \alpha}+\frac{1}{2}} \left(\frac{r_0^2}{8Dt}\right) \right]. 
\end{align}

The next step of the calculation consists of relating the variables imposed by Cauchy formula \mbox{$(M,\theta)$} and the polar coordinates \mbox{$(r_0,\varphi_0)$} in the wedge. This matching is given by the two relations (for details, see Appendix \ref{Geometry})
\begin{align}\label{geomrel}
&r_0=\dfrac{1}{\cos\theta} \sqrt{d^2+2 d M \sin\theta+M^2} \nonumber \\
&\varphi_0=\arccos\left( \frac{d+M \sin\theta}{\sqrt{M^2+2dM\sin\theta+d^2}}\right).
\end{align}
We introduce the non-dimensional variables 
\begin{align}\label{nondim}
&x=\frac{d}{\sqrt{Dt}} \nonumber \\
&u=\dfrac{M}{\sqrt{Dt}}.
\end{align}
Finally, a joint use of Eqs.~\eqref{defperim}, \eqref{survie} and \eqref{nondim} yields an expression of the mean perimeter of the convex hull at time $t$ for a Brownian motion starting at a distance $d$ from a reflecting wall, rescaled by $\sqrt{Dt}$ 
\begin{align}\label{LBessel}
&\tilde{L}(x) \equiv \left\langle \dfrac{L^{(d)}(t)}{\sqrt{Dt}} \right\rangle \nonumber \\
&  =2 \int_{-\pi/2}^{\pi/2}  \kern-0.3em d\theta \int_0^{+\infty} \kern-0.3em du \; \Bigg\{ 1 - \frac{\sqrt{x^2+2xu\sin\theta+u^2}}{\sqrt{\pi}\cos\theta} \nonumber \\ 
& \times \sum\limits_{m=0}^{\infty}\dfrac{\sin\left((2m+1)\frac{\pi}{\alpha}\arccos\left( \frac{x+u \sin\theta}{\sqrt{u^2+2xu\sin\theta+x^2}}\right)\right)}{2m+1}  \nonumber \\
&  \times e^{-\frac{x^2+2xu\sin\theta+u^2}{8 \cos^2\theta} }  \left[I_{\nu}\left(\frac{x^2+2xu\sin\theta+u^2}{8 \cos^2\theta}\right) \right. \nonumber \\
&  \qquad \qquad  \left. + \, I_{\nu+1}\left(\frac{x^2+2xu\sin\theta+u^2}{8 \cos^2\theta}\right) \right] \Bigg\}
\end{align}
with the index of Bessel function
\begin{equation}\label{defnu}
\nu=(2m+1)\frac{\pi}{2\alpha}-\frac{1}{2}.
\end{equation}
As expected, the mean rescaled perimeter is a scaling function, that only depends on the rescaled distance to the wall \mbox{$x=d/\sqrt{Dt}$}. It means in particular that the limits small distance $d$ and long time $t$ are equivalent. In what follows, we will always fix the observation time $t$ and study the impact of the distance $d$ on the mean rescaled perimeter.

Note that the formula for the scaling function ${\tilde L}(x)$ in Eq.~\eqref{LBessel}, albeit complicated, is exact and explicit for all rescaled distances $x\geqslant 0$. However, evaluating analytically the double integral and the infinite sum in Eq.~\eqref{LBessel}, or even plotting the function ${\tilde L}(x)$ numerically, is not easy. In subsection II.B, we show how to obtain an explicit expression of ${\tilde L}(0)$. In fact, even extracting the asymptotic behavior of the function ${\tilde L}(x)$, as \mbox{$x\to 0$}, turns out to be rather nontrivial, as we then demonstrate in subsection II.C. Finally, in subsection II.D, we provide a useful trick to evaluate numerically the right hand side of Eq.~\eqref{LBessel}, which we use to plot ${\tilde L}(x)$ numerically as a function of $x$ in Fig.~\ref{MRP}.

\subsection{Particular case of a Brownian motion starting from the wall}

We now show that in the important particular case where the Brownian motion starts from the reflecting wall, Eq.~\eqref{LBessel} assumes a very simple form. Indeed, geometric relations \eqref{geomrel} are significantly simpler in this case
\begin{align}
&r_0=\frac{M}{\cos \theta} \nonumber \\
& \varphi_0=\frac{\alpha}{2}
\end{align}
so the survival probability displayed in Eq.~\eqref{survie} can be rewritten in terms of the rescaled variables defined in Eq.~\eqref{nondim}
\begin{align}
&S^{(x)}(u,\theta)=\frac{u}{\sqrt{\pi} \cos\theta} \; e^{-\frac{u^2}{8 \cos^2\theta}} \sum\limits_{m=0}^{\infty} \frac{(-1)^m}{2m+1} \nonumber \\
&\quad  \times \left[ I_{\nu}\left(\frac{u^2}{8 \cos^2\theta}\right) +I_{\nu+1}\left(\frac{u^2}{8 \cos^2\theta}\right) \right].
\end{align}
Defining the mean rescaled maximum 
\begin{equation}
\tilde{\mathcal{M}}(\theta,x) \equiv \left\langle \frac{\mathcal{M}^{(d)}(\theta,t)}{\sqrt{Dt}} \right\rangle
\end{equation} 
and noticing that
\begin{equation}
\frac{u}{\sqrt{\pi}\cos\theta} \e{-\frac{u^2}{8\cos^2\theta}} \sum\limits_{m=0}^{+\infty}\frac{(-1)^m}{2m+1} \; 2 \; \frac{\e{\frac{u^2}{8\cos^2\theta}}}{\sqrt{2\pi\frac{u^2}{8\cos^2\theta}}}=1,
\end{equation}
we rewrite the $1$ of the integrand of Eq.~\eqref{defmax} and obtain
\begin{align}
& \tilde{\mathcal{M}}(\theta,0) =\frac{1}{\sqrt{\pi}\cos\theta} \sum\limits_{m=0}^{+\infty}\frac{(-1)^m}{2m+1} \int_0^{+\infty} du \; u \e{-\frac{u^2}{8\cos^2\theta}} \nonumber \\
& \times \left( \frac{2}{\sqrt{2\pi}} \frac{\e{\frac{u^2}{8\cos^2\theta}}}{\sqrt{\frac{u^2}{8\cos^2\theta}}} - I_{\nu}\left(\frac{u^2}{8 \cos^2\theta}\right) - I_{\nu+1}\left(\frac{u^2}{8 \cos^2\theta}\right) \right).
\end{align}
We next introduce the variable \mbox{$v=u^2/(8 \cos^2\theta)$}, so that
\begin{align}
& \tilde{\mathcal{M}}(\theta,0)  =\dfrac{4}{\sqrt{\pi}} \cos\theta  \sum\limits_{m=0}^{\infty} \dfrac{(-1)^m}{2m+1} \nonumber \\
& \quad \times \int_0^{+\infty}  dv \; e^{-v} \left( \sqrt{\dfrac{2}{\pi}} \dfrac{e^v}{\sqrt{v}}-I_{\nu}(v)-I_{\nu+1}(v)\right). 
\end{align}

We then introduce a parameter \mbox{$\beta \geq 1$} and define the following integral (that depends on the integer $m$ via $\nu$)
\begin{equation}
A(\beta,m) \equiv\int_0^{+\infty} dv \; e^{-\beta v} \left( \sqrt{\dfrac{2}{\pi}}\dfrac{e^v}{\sqrt{v}}-I_{\nu}(v)-I_{\nu+1}(v)\right).
\end{equation}
We recognize a Gamma function and the Laplace transform of the Bessel function \mbox{$\mathcal{L}[I_{\nu}]$}
\begin{equation}
A(\beta,m)=\sqrt{\dfrac{2}{\pi}}\dfrac{1}{\sqrt{\beta-1}} \; \Gamma\left(\dfrac{1}{2}\right)-\mathcal{L}[I_{\nu}](\beta)-\mathcal{L}[I_{\nu+1}](\beta),
\end{equation}
hence \cite{Prudnikov3}
\begin{align}
A(\beta,m)=&\sqrt{\dfrac{2}{\beta-1}}-\dfrac{(\beta+\sqrt{\beta^2-1})^{-\nu}}{\sqrt{\beta^2-1}} \nonumber \\
&-\dfrac{(\beta+\sqrt{\beta^2-1})^{-\nu-1}}{\sqrt{\beta^2-1}}.
\end{align}
We also define $B(\beta)$ such that \mbox{$\tilde{\mathcal{M}}(\theta,0)=B(1)$}
\begin{equation}
B(\beta)\equiv\frac{4}{\sqrt{\pi}} \cos\theta  \sum\limits_{m=0}^{\infty} \dfrac{(-1)^m}{2m+1} A(\beta,m).
\end{equation}\\

We then compute the following sums, setting \mbox{$a=\beta+\sqrt{\beta^2-1}$} and using Eq.~\eqref{defnu}
\begin{align}
&\sum\limits_{m=0}^{\infty} \frac{(-1)^m a^{-\frac{(2m+1)\pi}{2\alpha}+\frac{1}{2}}}{2m+1} \nonumber \\
& \quad = \frac{\pi}{2\alpha} \sqrt{a}\int_0^a dy \sum\limits_{m=0}^{\infty} (-1)^m y^{-\frac{(2m+1)\pi}{2\alpha}-1} \nonumber \\
&\quad = \frac{\pi}{2\alpha} \sqrt{a} \int_0^a dy \; \frac{y^{-\frac{\pi}{2\alpha}-1}}{1+y^{-\frac{\pi}{\alpha}}} = \sqrt{a} \arctan\left(a^{-\frac{\pi}{2\alpha}}\right) 
\end{align}
and similarly
\begin{equation}
\sum\limits_{m=0}^{\infty} \frac{(-1)^m a^{-\frac{(2m+1)\pi}{2\alpha}-\frac{1}{2}}}{2m+1} =\frac{1}{\sqrt{a}} \arctan\left(a^{-\frac{\pi}{2\alpha}}\right).
\end{equation}
This yields
\begin{equation}\label{B}
B(\beta)=\frac{4 \cos\theta}{\sqrt{\pi}} \left[ \frac{\pi}{4} \sqrt{\frac{2}{\beta-1}}  -\frac{\arctan\left(a^{-\frac{\pi}{2\alpha}}\right)}{\sqrt{\beta^2-1}}  \left(\frac{1}{\sqrt{a}} +\sqrt{a} \right) \right] 
\end{equation}
using that
\begin{equation}
\sum\limits_{m=0}^{\infty}\frac{(-1)^m}{2m+1}=\frac{\pi}{4}.
\end{equation}
As we need $B(1)$, we expand \mbox{$B(\beta)$}  for \mbox{$\beta=1+\epsilon$} with $\epsilon=o(1)$. We then have
\begin{align}
&a=\beta+\sqrt{\beta^2-1}=1+\epsilon+\sqrt{2\epsilon} \equi{\epsilon\to 0} 1+\sqrt{2\epsilon}, \nonumber \\
&\sqrt{a} \sim 1+\sqrt{\frac{\epsilon}{2}}, \nonumber \\
&\arctan\left( a^{-\frac{\pi}{2\alpha}} \right) \sim \arctan \left( 1-\dfrac{\pi}{\alpha} \sqrt{\frac{\epsilon}{2}} \right) \sim \frac{\pi}{4} \left( 1- \frac{\sqrt{2\epsilon}}{\alpha} \right). \nonumber
\end{align}
Plugging these developments into Eq.~\eqref{B} and taking \mbox{$\epsilon=0$}, we finally obtain the mean rescaled maximum in the direction $\theta$ starting from the reflecting wall
\begin{equation}
 \tilde{\mathcal{M}}(\theta,0)= 2 \sqrt{\pi} \frac{\cos\theta}{\pi-2\theta}.
\end{equation}
Let us consider several remarkable directions $\theta$. First, for \mbox{$\theta=0$}, that is to say in the direction parallel to the reflecting wall, one finds
\begin{equation}
 \tilde{\mathcal{M}}(0,0)=\frac{2}{\sqrt{\pi}} \simeq 1.128.
\end{equation}
Hence, the mean span in the direction parallel to the wall is
\begin{equation}\label{span0}
S(0)= \tilde{\mathcal{M}}(0,0)+ \tilde{\mathcal{M}}(\pi,0)=2 \; \tilde{\mathcal{M}}(0,0)= \frac{4}{\sqrt{\pi}}
\end{equation}
which is exactly the result obtained in the non-confined case. Indeed, the potential reflections on the wall do not affect the Brownian motion in the parallel direction.

For \mbox{$\theta=-\pi/2$}, that is to say in the direction orthogonal to the wall towards it, we find, as expected, that the mean extension is zero, as the Brownian motion cannot go farther in this direction, blocked by the wall.

For \mbox{$\theta=\pi/2$}, orthogonally to the wall away from it, we find a result higher than in the non-confined case. Indeed, the wall pushes the trajectories farther in this direction
\begin{equation}
\tilde{\mathcal{M}}\left(\frac{\pi}{2},0 \right) = \sqrt{\pi}\simeq 1.772.
\end{equation}

Eventually, it is straightforward to obtain the mean rescaled perimeter of the convex hull by integrating over the angle $\theta$. We finally obtain the simple result
\begin{equation}
\tilde{L}(0)=2 \int_{-\frac{\pi}{2}}^{\frac{\pi}{2} } d\theta \; 2 \sqrt{\pi} \frac{\cos\theta}{\pi-2\theta}= 2 \sqrt{\pi} \, \mathrm{Si}(\pi) \simeq 6.565.
\end{equation}
Note that this value is lower than the mean rescaled perimeter of the convex hull in the absence of confinement (\mbox{$4\sqrt{\pi} \simeq 7.090$}).

\subsection{Case of a Brownian motion starting near the wall}

We now focus on the case where the starting point is close to the wall, i.e. \mbox{$d \ll \sqrt{Dt}$}, or in terms of the rescaled distance $x$, \mbox{$x \ll 1$}. 

To obtain a development at small $x$ of the mean rescaled perimeter, we seek an expression of the mean maximum \mbox{$\tilde{\mathcal{M}}(\theta,x)$} as a power series in $x$. As we will see later on, the development of \mbox{$\tilde{\mathcal{M}}(\theta,x)$} depends on the sign of $\theta$. We first derive the development of the maximum for positive angles $\theta$ using a small $x$ development of the survival probability and straightforwardly computing Laplace transforms. The development of the maximum for negative $\theta$, presented afterwards, relies in part on the same derivation as for the case \mbox{$\theta>0$}, but requires a slight modification.
 The zero-order term is given by the \mbox{$d=0$} case, addressed in the previous subsection, so we will only focus on higher order terms. 

We remind that the expression of the survival probability in terms of \mbox{$x=d/\sqrt{Dt}$} and \mbox{$u=M/\sqrt{Dt}$} is
\begin{align}
& S^{(x)}(t|u,\theta)= \frac{\sqrt{x^2+2xu\sin\theta+u^2}}{\sqrt{\pi}\cos\theta}   e^{-\frac{x^2+2xu\sin\theta+u^2}{8 \cos^2\theta}}  \times \nonumber \\
&  \sum\limits_{m=0}^{\infty}\frac{\sin\left((2m+1)\frac{\pi}{\alpha}\arccos\left( \frac{x+u \sin\theta}{\sqrt{u^2+2xu\sin\theta+x^2}}\right)\right)}{2m+1}  \times \nonumber \\
&  \left[I_{\nu}\left(\frac{x^2+2xu\sin\theta+u^2}{8 \cos^2\theta}\right) +I_{\nu+1}\left(\frac{x^2+2xu\sin\theta+u^2}{8 \cos^2\theta}\right) \right]
\end{align}
with \mbox{$\alpha=\pi-2\theta$} and \mbox{$\nu=(2m+1)\pi/(2\alpha)-1/2$}.

\subsubsection{Development of $\tilde{\mathcal{M}}(\theta,x)$ for $\theta>0$}\label{thetapos}

We first focus on the development of \mbox{$\tilde{\mathcal{M}}(\theta,x)$} for $\theta$ in the range \mbox{$[0,\pi/2]$}. The derivation given below does not hold for $\theta$ in \mbox{$[-\pi/2,0]$} as divergences arise. We will deal with this latter case afterwards. 

We start by developing the survival probability at small $x$. We can write
\begin{align}
&\sqrt{u^2+2xu\sin\theta+x^2}  \sim u + x\sin\theta +\frac{x^2}{2u} \cos^2\theta, \\
&e^{-\frac{x^2+2xu\sin\theta}{8 \cos^2\theta}}  \sim 1-\frac{xu\sin\theta}{4 \cos^2\theta}-\frac{x^2}{8 \cos^2\theta}+\frac{x^2u^2\sin^2\theta}{32 \cos^4\theta}, \nonumber \\
& \\
&I_{\nu}\left(\frac{u^2+2xu\sin\theta+x^2}{8 \cos^2\theta}\right)  \sim I_{\nu}\left(\frac{u^2}{8 \cos^2\theta}\right) \nonumber \\
& \qquad \qquad + \left( \frac{xu\sin\theta}{4\cos^2\theta} + \frac{x^2}{8 \cos^2\theta} \right)  I'_{\nu}\left(\frac{u^2}{8 \cos^2\theta}\right) \nonumber \\
& \qquad \qquad + \frac{x^2u^2\sin^2\theta}{32 \cos^4\theta}  I''_{\nu}\left(\frac{u^2}{8 \cos^2\theta}\right)
\end{align}
and
\begin{align}
\arccos\left( \frac{x+u \sin\theta}{\sqrt{u^2+2xu\sin\theta+x^2}}\right) & \sim \arccos \left(\sin\theta +\frac{x}{u} \cos^2\theta\right) \nonumber \\
& \sim \frac{\pi}{2}-\theta -\frac{x}{u} \cos\theta,
\end{align}
yielding
\begin{align}
&\sin\left((2m+1)\frac{\pi}{\alpha}\arccos\left( \frac{x+u \sin\theta}{\sqrt{u^2+2xu\sin\theta+x^2}}\right)\right) \nonumber \\
& \qquad  \sim (-1)^m \cos\left((2m+1)\pi \frac{x}{u}\frac{\cos\theta}{\pi-2\theta}\right) \nonumber \\
& \qquad \sim (-1)^m \left( 1-\frac{(2m+1)^2}{2} \frac{\pi^2}{(\pi-2\theta)^2} \frac{x^2}{u^2} \cos^2\theta \right). \nonumber \\
\end{align}
The derivative of Bessel functions can be expressed in three different ways \citep{Abramowitz}. We successively use the following expressions
\begin{align}
 I'_{\nu}(x) &=\frac{I_{\nu-1}(x)+I_{\nu+1}(x)}{2} \label{Besselsomme}\\
& =I_{\nu-1}(x)-\frac{\nu}{x} I_{\nu}(x) \label{Besselmoins}\\
&= I_{\nu+1}(x)+\frac{\nu}{x	} I_{\nu}(x). \label{Besselplus}
\end{align}

\paragraph{First-order term.}

We extract the expression of the first-order term in $x$ \mbox{$\Delta_1(\theta)$} of the maximum from the previous development
\begin{align}
&\Delta_1(\theta)=-\sum\limits_{m=0}^{+\infty} \frac{(-1)^m}{2m+1} \int_0^{+\infty} du \frac{e^{-\frac{u^2}{8\cos^2\theta}}}{\sqrt{\pi}\cos\theta} \nonumber \\
&\times \left[ \left(\sin\theta-\frac{u^2\sin\theta}{4\cos^2\theta} \right) \left( I_{\nu} \left(\frac{u^2}{8\cos^2\theta}\right) +I_{\nu+1} \left(\frac{u^2}{8\cos^2\theta} \right)\right) \right. \nonumber \\
& \left.  \qquad +\frac{u^2 \sin\theta}{4 \cos^2\theta} \left( I'_{\nu} \left(\frac{u^2}{8\cos^2\theta} \right) +I'_{\nu+1} \left(\frac{u^2}{8\cos^2\theta} \right)\right) \right] \nonumber \\
\end{align}
Using  \eqref{Besselplus} to express \mbox{$I'_{\nu} \left(u^2/8\cos^2\theta\right)$} and  \eqref{Besselmoins} for \mbox{$I'_{\nu+1} \left(u^2/8\cos^2\theta \right)$}, we get
\begin{align}
&I'_{\nu} \left(\frac{u^2}{8\cos^2\theta} \right) +I'_{\nu+1} \left(\frac{u^2}{8\cos^2\theta} \right) =\nonumber \\
 &\quad \left(\frac{8 \cos^2\theta}{u^2} \nu +1 \right) I_{\nu}\left(\frac{u^2}{8 \cos^2\theta} \right) \nonumber \\
& \qquad +\left( 1- \frac{(\nu+1) 8\cos^2\theta}{u^2} \right) I_{\nu+1}\left( \frac{u^2}{8\cos^2\theta}\right) 
\end{align}
which simplifies the expression of \mbox{$\Delta_1(\theta)$}
\begin{align}
&\Delta_1(\theta)=-\sum\limits_{m=0}^{+\infty} \frac{(-1)^m}{2m+1} \int_0^{+\infty} du \frac{e^{-\frac{u^2}{8\cos^2\theta}}}{\sqrt{\pi}\cos\theta} \nonumber \\
&\times \sin\theta \; (1+2\nu) \left(I_{\nu}\left(\frac{u^2}{8\cos^2\theta} \right) -I_{\nu+1}\left(\frac{u^2}{8\cos^2\theta} \right) \right), \nonumber\\
\end{align}
and setting \mbox{$y=u^2/8\cos^2\theta$},
\begin{align}
\Delta_1(\theta)=&-\sqrt{\frac{2}{\pi}}  \sum\limits_{m=0}^{+\infty} \frac{(-1)^m}{2m+1} \; (1+2\nu)  \sin\theta    \nonumber \\
& \qquad \times \int_0^{+\infty} \frac{dy}{\sqrt{y}} e^{-y} \left(I_{\nu}(y) -I_{\nu+1}(y) \right).
\end{align}

Separating the two parts of the integrand would lead to a divergence, but as it is written in Eq.~\eqref{Delta1iny}, the integral over $y$ is finite as it involves a difference of Bessel functions. We use the same trick as previously to calculate this integral: we introduce a parameter \mbox{$\beta\geqslant 1$} and compute the limit \mbox{$\beta \to 1$}. We set
\begin{equation}
D(\beta,m) \equiv \int_0^{+\infty} \frac{dy}{\sqrt{y}} e^{-\beta y} \left(I_{\nu}(y) -I_{\nu+1}(y) \right)
\end{equation}
such that
\begin{equation}\label{Delta1iny}
\Delta_1(\theta)=-\sqrt{\frac{2}{\pi}}  \sum\limits_{m=0}^{+\infty} \frac{(-1)^m}{2m+1} \; (1+2\nu)  \sin\theta \; D(1,m),
\end{equation}
and using \cite{Prudnikov2},
 \begin{equation}
D(\beta,m)= \sqrt{\frac{2}{\pi}} \left( Q_{\nu-\frac{1}{2}}(\beta)-Q_{\nu+\frac{1}{2}}(\beta) \right)
\end{equation}
where \mbox{$Q_{\nu-\frac{1}{2}}(\beta)$} is a Legendre function of the second kind, defined by
\begin{align}
&Q_{\nu-\frac{1}{2}}(\beta)=\frac{1}{2^{\nu+\frac{1}{2}}} \sqrt{\pi} \; \frac{\Gamma(\nu+\frac{1}{2})}{\Gamma(\nu+1)} \frac{1}{\beta^{\nu+\frac{1}{2}}}  \nonumber \\
& \times \, _2F_1\left(\frac{\nu}{2}+\frac{3}{4},\frac{\nu}{2}+\frac{1}{4};\nu+1; \frac{1}{\beta^2} \right)
\end{align}
with $_2F_1$ a hypergeometric function. Writing \mbox{$Q_{\nu+\frac{1}{2}}(\beta)$} with the same formula and using the relation
\begin{equation}\label{relationGamma}
\Gamma(x+1)=x \; \Gamma(x),
\end{equation}
we obtain
\begin{align}
&D(\beta,m)=\frac{1}{2^{\nu}} \frac{\Gamma(\nu+\frac{1}{2})}{\Gamma(\nu+1)} \; \frac{1}{\beta^{\nu+\frac{1}{2}}} \nonumber \\
& \times  \left[ _2F_1 \left( \frac{\nu}{2}+\frac{3}{4},\frac{\nu}{2}+\frac{1}{4};\nu+1; \frac{1}{\beta^2} \right) \right. \nonumber \\
& \qquad \left. - \frac{1}{2 \beta} \frac{\nu+\frac{1}{2}}{\nu+1} \,_2F_1 \left( \frac{\nu}{2}+\frac{5}{4},\frac{\nu}{2}+\frac{3}{4};\nu+2; \frac{1}{\beta^2} \right) \right]. 
\end{align}
Hypergeometric functions of the type \mbox{$_2F_1(a,b;a+b;z)$} can be expressed
\begin{align}
& _2F_1 \left(  \frac{\nu}{2} + \frac{3}{4},\frac{\nu}{2}+\frac{1}{4};\nu+1; \frac{1}{\beta^2} \right)= \frac{\Gamma(\nu+1)}{\Gamma(\frac{\nu}{2}+\frac{3}{4}) \Gamma(\frac{\nu}{2}+\frac{1}{4})}   \nonumber \\
& \quad  \times \sum\limits_{n=0}^{+\infty}\frac{\left( \frac{\nu}{2}+\frac{3}{4} \right)_n \left( \frac{\nu}{2}+\frac{1}{4} \right)_n}{(n!)^2} \left( 1-\frac{1}{\beta^2}\right)^n \nonumber \\
&\quad  \times \left[2 \psi(n+1)-\psi\left(\frac{\nu}{2}+\frac{3}{4}+n\right)-\psi\left(\frac{\nu}{2}+\frac{1}{4}+n\right) \right.\nonumber \\
& \quad \qquad \left. -\ln\left(1-\frac{1}{\beta^2} \right) \right]  
\end{align}
with \mbox{$(a)_n=\Gamma(a+n)/\Gamma(a)$} the Pochhammer symbol, and \mbox{$\psi(x)=\Gamma'(x)/\Gamma(x)$} the digamma function. As we take the limit \mbox{$\beta \to 1$}, the term \mbox{$n=0$} is the only non-zero term in the sum. We then have
\begin{align}
&D(\beta,m) \equi{\beta \to 1} \frac{1}{2^{\nu}} \frac{\Gamma(\nu+\frac{1}{2})}{\Gamma(\nu+1)} \; \frac{1}{\beta^{\nu+\frac{1}{2}}}  \nonumber \\
& \times \left[ \frac{\Gamma(\nu+1)}{\Gamma(\frac{\nu}{2}+\frac{3}{4}) \Gamma(\frac{\nu}{2}+\frac{1}{4})} \left( 2 \psi(1)-\psi\left(\frac{\nu}{2}+\frac{3}{4}\right) \right. \right. \nonumber \\
& \qquad \qquad \left. -\psi\left(\frac{\nu}{2}+\frac{1}{4}\right)  -\ln\left(1-\frac{1}{\beta^2} \right) \right)  \nonumber \\
& -\frac{1}{2\beta} \; \frac{\nu+\frac{1}{2}}{\nu+1} \; \frac{\Gamma(\nu+2)}{\Gamma(\frac{\nu}{2}+\frac{5}{4}) \Gamma(\frac{\nu}{2}+\frac{3}{4})} \times \nonumber \\
&  \left. \left( 2 \psi(1) -\psi\left(\frac{\nu}{2}+\frac{5}{4}\right) -\psi\left(\frac{\nu}{2}+\frac{3}{4}\right)  -\ln\left(1-\frac{1}{\beta^2} \right) \right) \right]. \nonumber \\
\end{align}
Using the relation (\ref{relationGamma}) and
\begin{equation}
\psi(x+1)=\psi(x)+\frac{1}{x},
\end{equation}
we obtain
\begin{equation}
D(\beta,m) \underset{\beta\to 1}{\longrightarrow} \frac{1}{2^{\nu}} \; \frac{\Gamma(\nu+\frac{1}{2})}{\Gamma(\frac{\nu}{2}+\frac{3}{4}) \Gamma(\frac{\nu}{2}+\frac{1}{4})}  \; \frac{1}{\frac{\nu}{2}+\frac{1}{4}}. 
\end{equation}
Making use of the following property
\begin{equation}
\Gamma(2z)=\frac{1}{\sqrt{2\pi}} 2^{2z-\frac{1}{2}} \; \Gamma(z) \;\Gamma(z+\frac{1}{2}),
\end{equation}
we eventually obtain a very simple expression for \mbox{$D(1,m)$}
\begin{equation}
D(1,m)=\sqrt{\frac{2}{\pi}} \frac{2}{1+2\nu}.
\end{equation}
Plugging this expression into Eq.~\eqref{Delta1iny}, the first-order term of \mbox{$\tilde{\mathcal{M}}(\theta,x)$ for $\theta$} finally is 
\begin{equation}
\Delta_1=-\sin \theta.
\end{equation}

\paragraph{Second-order term.}
Applying again the change in variables \mbox{$y=u^2/(8\cos^2\theta)$}, the second-order term is
\begin{align}
&\Delta_2(\theta) =-\frac{1}{\sqrt{\pi}} \sum\limits_{m=0}^{+\infty} \frac{(-1)^m}{2m+1} \int_0^{+\infty} \sqrt{2} \; dy \; e^{-y} \times \nonumber \\
&\left[ \frac{\cos\theta}{4\sqrt{2}y} \left(I_{\nu}(y)+I_{\nu+1}(y) \right) \left(1-\frac{\pi^2(2m+1)^2}{(\pi-2\theta)^2} \right) \right. \nonumber \\
& + \frac{1+2\sin^2\theta}{4\sqrt{2}\cos\theta} \left(I_{\nu-1}-I_{\nu}-I_{\nu+1}+I_{\nu+2} \right)(y) \nonumber \\
& +  \frac{\sin^2\theta}{4 \sqrt{2} \cos\theta} \; y \; \times \nonumber \\
& \left(I_{\nu-2}-3I_{\nu-1}+2I_{\nu}+2I_{\nu+1}-3I_{\nu+2}+I_{\nu+3} \right)(y) \Big]. 
\end{align}
Note that this equation is valid only if all \mbox{$\nu>0$}, which is true for \mbox{$\theta>0$} but wrong otherwise (since \mbox{$\nu(0)=\theta/\alpha$}), as pointed out previously. In the latter case, the coefficient \mbox{$\Delta_2(\theta)$} under this form is infinite because of the term \mbox{$I_{\nu(0)}/y$} is not integrable in 0. 

We again introduce a parameter \mbox{$\beta\geqslant 1$} to calculate the integral and take the limit \mbox{$\beta\to 1$} afterwards. We define
\begin{align}
&E(\beta,m) \equiv \int_0^{+\infty} dy \; e^{-\beta y}  \times \nonumber \\
&\left[ \frac{\cos\theta}{4\sqrt{2}y} \left(I_{\nu}(y)+I_{\nu+1}(y) \right) \left(1-\frac{\pi^2(2m+1)^2}{(\pi-2\theta)^2} \right) \right. \nonumber \\
&+ \frac{1+2\sin^2\theta}{4\sqrt{2}\cos\theta} \left(I_{\nu-1}-I_{\nu}-I_{\nu+1}+I_{\nu+2} \right)(y) \nonumber \\
& + y \frac{\sin^2\theta}{4 \sqrt{2} \cos\theta} \times \nonumber \\
&  \left(I_{\nu-2}-3I_{\nu-1}+2I_{\nu}+2I_{\nu+1}-3I_{\nu+2}+I_{\nu+3} \right)(y) \Big] 
\end{align}
which is equivalent to
\begin{align}
&E(\beta,m)= \frac{\cos\theta}{4\sqrt{2}}  \left(1-\frac{\pi^2(2m+1)^2}{(\pi-2\theta)^2} \right) \mathcal{L}\left[\frac{I_{\nu}(y)+I_{\nu+1}(y)}{y} \right](\beta)  \nonumber \\
&\quad + \frac{(1+2\sin^2\theta)}{4\sqrt{2}\cos\theta} \mathcal{L}\left[I_{\nu-1}-I_{\nu}-I_{\nu+1}+I_{\nu+2}\right](\beta) \nonumber \\
&  \quad +\frac{\sin^2\theta}{4\sqrt{2}\cos\theta} \mathcal{L} \left[y (I_{\nu-2}-3I_{\nu-1}+2I_{\nu}+2I_{\nu+1} \right.\nonumber \\
& \qquad \qquad \qquad \qquad \left.-3I_{\nu+2}+I_{\nu+3})(y)\right](\beta),
\end{align}
such that
\begin{equation}
\Delta_2(\theta) =-\sqrt{\frac{2}{\pi}} \sum\limits_{m=0}^{+\infty} \frac{(-1)^m}{2m+1} \; E(1,m).
\end{equation}
These Laplace transforms have known expressions \cite{Prudnikov3}
\begin{align}
&\mathcal{L}\left[\frac{I_{\nu}(y)}{y} \right](\beta) = \frac{1}{\nu} (\beta+\sqrt{\beta^2-1})^{-\nu} \underset{\beta\to 1}{\longrightarrow}  \frac{1}{\nu} \label{lapl1sury}\\
&\mathcal{L}\left[I_{\nu}(y)\right](\beta) = \frac{(\beta+\sqrt{\beta^2-1})^{-\nu}}{\sqrt{\beta^2-1}} \nonumber \\
&  \qquad \qquad \underset{\epsilon\to0}{\underset{\beta=1+\epsilon}{=}} \frac{1}{\sqrt{2\epsilon}} -\nu +o(1)  \\
&\mathcal{L}\left[y \; I_{\nu}(y)\right](\beta) = \frac{\beta+\nu\sqrt{\beta^2-1}}{(\beta+\sqrt{\beta^2-1})^{\nu}(\beta^2-1)^{3/2}} \nonumber \\
& \underset{\epsilon\to0}{\underset{\beta=1+\epsilon}{=}}\frac{1}{2\epsilon^{3/2}} \left[ 1+\epsilon \left( \frac{1}{4}-\nu^2\right) +\sqrt{2}\epsilon^{3/2} ( \nu^2+\nu^3) \right] +o(1). 
\end{align}
Carrying out simple algebra, we get 
\begin{align}
&\mathcal{L}\left[\frac{I_{\nu}(y)+I_{\nu+1}(y)}{y} \right](\beta) \underset{\beta\to1}{\longrightarrow} \frac{1}{\nu}+\frac{1}{\nu+1} \nonumber \\
&\qquad \qquad =\frac{(2m+1)\pi}{\pi-2\theta} \frac{4}{\left(\frac{(2m+1)^2\pi^2}{(\pi-2\theta)^2}-1 \right)}  \\
 & \mathcal{L}\left[I_{\nu-1}-I_{\nu}-I_{\nu+1}+I_{\nu+2}\right](\beta)  \underset{\beta\to1}{\longrightarrow}0 \\
& \mathcal{L} \left[y (I_{\nu-2}-3I_{\nu-1}+2I_{\nu}+2I_{\nu+1} \right. \nonumber \\
& \qquad \qquad \qquad \qquad \left. -3I_{\nu+2}+I_{\nu+3})(y)\right](\beta) \underset{\beta\to1}{\longrightarrow}0. 
\end{align}
The second-order term is finally
\begin{equation}
\Delta_2(\theta)=\frac{\sqrt{\pi}}{2} \frac{\cos\theta}{\pi-2\theta}.
\end{equation}

The development of the mean rescaled maximum in the direction \mbox{$\theta>0$} \mbox{$\tilde{\mathcal{M}}(\theta,x) $} up to order 2 is then
\begin{equation}\label{Mposdev}
\tilde{\mathcal{M}}(\theta,x) =\frac{2\sqrt{\pi}\cos\theta}{\pi-2\theta} -\sin\theta \, x + \frac{\sqrt{\pi}}{2} \frac{\cos \theta}{\pi-2\theta} x^2 +o(x^2).
\end{equation}

\subsubsection{Development of $\tilde{\mathcal{M}}(\theta,x)$ for $\theta<0$}

We now focus on the development of \mbox{$\tilde{\mathcal{M}}(\theta,x)$} for $\theta$ in the range \mbox{$[-\pi/2,0]$}. To bypass the problem of convergence raised in the previous paragraph, we isolate the term that yields the divergence, namely the term \mbox{$m=0$} of the sum, and only the part in \mbox{$I_{\nu}$} of this term. The other terms (the part \mbox{$I_{\nu+1}$} of the term \mbox{$m=0$}, and the terms \mbox{$m>0$} of the sum) can be developed until order 2 following the same lines as previously. We therefore separate \mbox{$ \tilde{M}(\theta,x)$} into three parts
\begin{equation}\label{defT}
\tilde{M}(\theta,x)=T_0(\theta,x)+T'_0(\theta,x)+T_1(\theta,x)
\end{equation}
with 
\begin{align}
&T_0(\theta,x)=\int_0^{+\infty} \kern-1em du \left[ \frac{2}{\pi} -\frac{\sqrt{x^2+2xu\sin\theta+u^2}}{\sqrt{\pi}\cos\theta}  \times  \right. \nonumber \\
&  \quad  \cos\left(\frac{\pi}{\alpha}\arccos\left( \frac{u+x \sin\theta}{\sqrt{u^2+2xu\sin\theta+x^2}}\right)\right) \times \nonumber \\
& \quad  \left. e^{-\frac{x^2+2xu\sin\theta+u^2}{8 \cos^2\theta}}  I_{\nu(0)}\left(\frac{x^2+2xu\sin\theta+u^2}{8 \cos^2\theta}\right) \right], 
\end{align}
\begin{align}
&T'_0(\theta,x)=\int_0^{+\infty} \kern-1em du \left[ \frac{2}{\pi} -\frac{\sqrt{x^2+2xu\sin\theta+u^2}}{\sqrt{\pi}\cos\theta}  \times   \right. \nonumber \\
&  \quad \cos\left(\frac{\pi}{\alpha}\arccos\left( \frac{u+x \sin\theta}{\sqrt{u^2+2xu\sin\theta+x^2}}\right)\right) \times \nonumber \\
&  \quad  e^{-\frac{x^2+2xu\sin\theta+u^2}{8 \cos^2\theta}} \left. I_{\nu(0)+1}\left(\frac{x^2+2xu\sin\theta+u^2}{8 \cos^2\theta}\right) \right] 
\end{align}
and
\begin{align}
&T_1(\theta,x)=\sum\limits_{m=1}^{+\infty} \frac{(-1)^m}{2m+1}  \int_0^{+\infty} \kern-1em du \kern-0.3em \left\{ \frac{4}{\pi} -\frac{\sqrt{x^2+2xu\sin\theta+u^2}}{\sqrt{\pi}\cos\theta}  \right.\nonumber \\
&\times  \cos\left(  (2m+1)\frac{\pi}{\alpha}\arccos\left( \frac{u+x \sin\theta}{\sqrt{u^2+2xu\sin\theta+x^2}}\right) \right) \nonumber \\
& \times \left. e^{-\frac{x^2+2xu\sin\theta+u^2}{8 \cos^2\theta}} \left[ (I_{\nu}+\, I_{\nu+1}) \left(\frac{x^2+2xu\sin\theta+u^2}{8 \cos^2\theta}\right)  \right] \right\}. 
\end{align}

To compute the term $T_0$, which would yield a divergence if we were using the previous derivation, it is convenient to transfer the $x$ dependence to the Bessel function. We thus change the variables
\begin{align}
&z=\frac{u^2+2xu\sin\theta+x^2}{x^2\cos^2\theta} \nonumber \\
&du=\begin{cases} -\frac{x\cos\theta}{2\sqrt{z-1}} dz &\textrm{if}\quad  u\leqslant -x\sin\theta \nonumber \\
\frac{x\cos\theta}{2\sqrt{z-1}} dz &\textrm{if} \quad u\geqslant -x\sin\theta \nonumber \\
\end{cases}
\end{align}
and obtain
\begin{align}
&\kern-1em T_0(\theta,x)=\int_1^{\frac{1}{\cos^2\theta}} dz \frac{x\cos\theta}{2\sqrt{z-1}} \left[\frac{2}{\pi}-\frac{x}{\sqrt{\pi}} \sqrt{z} \right. \nonumber \\
& \left. \times  \cos\left(\frac{\pi}{\alpha}\arccos\left(-\frac{\sqrt{z-1}}{\sqrt{z}}\right) \right) \e{-\frac{x^2 z}{8}} \textrm{I}_{\nu(0)}\left(\frac{x^2 z}{8}\right) \right] \nonumber \\
&  +\int_1^{+\infty} dz \frac{x\cos\theta}{2\sqrt{z-1}} \left[\frac{2}{\pi}-\frac{x}{\sqrt{\pi}} \sqrt{z} \right. \nonumber \\
& \left. \times  \cos\left(\frac{\pi}{\alpha}\arccos\left(\frac{\sqrt{z-1}}{\sqrt{z}}\right) \right) \e{-\frac{x^2 z}{8}} \textrm{I}_{\nu(0)}\left(\frac{x^2 z}{8}\right) \right] \nonumber \\
&\equiv \mathcal{I}_1+\mathcal{I}_2.
\end{align}
The first integral \mbox{$\mathcal{I}_1$} can be written as
\begin{align}
&\mathcal{I}_1=-\frac{2x}{\pi}\sin\theta -\frac{x^2 \cos\theta}{2\sqrt{\pi}} \int_1^{\frac{1}{\cos^2\theta}} dz \sqrt{\frac{z}{z-1}} \nonumber \\
& \times \cos\left(\frac{\pi}{\alpha}\arccos\left(-\frac{\sqrt{z-1}}{\sqrt{z}}\right) \right) \e{-\frac{x^2 z}{8}} \textrm{I}_{\nu(0)}\left(\frac{x^2 z}{8}\right). 
\end{align}
For the second integral, we cannot split the integral as we did for \mbox{$\mathcal{I}_1$} since the upper limit is infinite. We have to make the order 0 in $x$ of this term appear in \mbox{$\mathcal{I}_2$}, given by
\begin{align}
&\int_0^{+\infty} dz \frac{x\cos\theta}{2\sqrt{z}} \left[ \frac{2}{\pi} -\frac{x}{\sqrt{\pi}} \sqrt{z} \e{-\frac{x^2 z}{8}} \textrm{I}_{\nu(0)}\left(\frac{x^2 z}{8}\right) \right] \nonumber \\
& \quad = 4 \, \nu(0) \frac{\cos\theta}{\sqrt{\pi}}.
\end{align}
It allows to split the integral
\begin{align}
&\mathcal{I}_2=4 \, \nu(0) \frac{\cos\theta}{\sqrt{\pi}}+\int_1^{+\infty} dz \frac{x\cos\theta}{\pi} \left(\frac{1}{\sqrt{z-1}}-\frac{1}{\sqrt{z}} \right) \nonumber \\
& -\frac{x^2\cos\theta}{2\sqrt{\pi}} \int_1^{+\infty} \kern-0.7em dz \kern-0.3em \left[\sqrt{\frac{z}{z-1}} \cos\left(\frac{\pi}{\alpha} \arccos\left(\sqrt{\frac{z-1}{z}}\right) \kern-0.3em \right)-1 \right] \nonumber \\
& \qquad \qquad \qquad \times \e{-\frac{x^2 z}{8}} \textrm{I}_{\nu(0)}\left(\frac{x^2 z}{8}\right) \nonumber \\
& -\int_0^1 dz \frac{x\cos\theta}{\pi \sqrt{z}} +\frac{x^2\cos\theta}{2\sqrt{\pi}} \int_0^1 dz \e{-\frac{x^2 z}{8}} \textrm{I}_{\nu(0)}\left(\frac{x^2 z}{8}\right).
\end{align}
This finally yields 
\begin{equation}
T_0(\theta,x)=\frac{4\theta\cos\theta}{\sqrt{\pi}(\pi-2\theta)}-\frac{2x}{\pi}\sin\theta +C(x,\theta)
\end{equation}
with
\begin{align}
&C(x,\theta)\equiv\frac{x^2\cos\theta}{2\sqrt{\pi}} \left\{\int_0^1 dz \e{-\frac{x^2 z}{8}} \textrm{I}_{\nu(0)}\left(\frac{x^2 z}{8}\right) \right. \nonumber \\
& - \int_1^{\frac{1}{\cos^2\theta}} \kern-0.5em dz \sqrt{\frac{z}{z-1}}\cos\left(\frac{\pi}{\alpha}\arccos\left(-\frac{\sqrt{z-1}}{\sqrt{z}}\right) \right) \nonumber \\
& \qquad \qquad \times \e{-\frac{x^2 z}{8}} \textrm{I}_{\nu(0)}\left(\frac{x^2 z}{8}\right) \nonumber \\
& -\int_1^{+\infty} \kern-0.5em dz \left[\sqrt{\frac{z}{z-1}} \cos\left(\frac{\pi}{\alpha} \arccos\left(\sqrt{\frac{z-1}{z}}\right) \right)-1 \right] \nonumber \\
& \left. \qquad \qquad \times \e{-\frac{x^2 z}{8}} \textrm{I}_{\nu(0)}\left(\frac{x^2 z}{8}\right) \right\}.
\end{align}

We use the development of the integrand for small $x$
\begin{equation}\label{devbessel}
\e{-\frac{x^2 z}{8}} \textrm{I}_{\nu(0)}\left(\frac{x^2 z}{8}\right) \equi{x\to0} \frac{x^{2\nu(0)}z^{\nu(0)}}{16^{\nu(0)} \Gamma(1+\nu(0))} +O(x^{2+2\nu(0)})
\end{equation}
and obtain
\begin{equation}
C(x,\theta) \equi{x\to0} C(\theta) x^{2+2\nu(0)} + C_2(x,\theta)+o(x^2)
\end{equation}
where
\begin{align}\label{defC}
&C(\theta)\equiv\frac{\cos\theta}{2^{4\nu(0)+1} \sqrt{\pi} \; \Gamma(1+\nu(0))}   \Bigg\{ \frac{1}{1+\nu(0)} \nonumber \\
& - \int_1^{\frac{1}{\cos^2\theta}} \kern-0.5em dz \sqrt{\frac{z}{z-1}}\cos\left(\frac{\pi}{\alpha}\arccos\left(-\frac{\sqrt{z-1}}{\sqrt{z}}\right) \right) z^{\nu(0)} \nonumber \\
&  -\int_1^{+\infty} \kern-1em dz \left[\sqrt{\frac{z}{z-1}} \cos\left(\frac{\pi}{\alpha} \arccos\left(\sqrt{\frac{z-1}{z}}\right) \right)-1 \right] z^{\nu(0)} \Bigg\} \nonumber \\
\end{align}
and 
\begin{align} \label{defC2}
&C_2(x,\theta) \equiv  -\frac{\cos\theta x^2}{2\sqrt{\pi}} \nonumber \\
& \times \int_1^{+\infty} \kern-0.8em dz \left[ \sqrt{\frac{z}{z-1}} \cos\left( \frac{\pi}{\alpha} \arccos\left( \sqrt{\frac{z-1}{z}} \right) \right) -1 \right] \nonumber \\
& \times \left[ e^{-\frac{x^2z}{8}} I_{\nu}\left(\frac{x^2 z}{8} \right) -\frac{x^{2\nu(0)} z^{\nu(0)}}{16^{\nu(0)} \Gamma(1+\nu(0))} \right].
\end{align}
If we introduce the variable \mbox{$u=x^2z/8$} in \mbox{$C_2(x,\theta)$} and use the following development
\begin{align}
&\frac{1}{\sqrt{1-\frac{x^2}{8u}}} \cos\left(\frac{\pi}{\alpha}\arccos\left(\sqrt{1-\frac{x^2}{8u}}\right) \right) \nonumber \\
&\qquad \equi{x\to0} 1+\left(1-\frac{\pi^2}{\alpha^2}\right) \frac{x^2}{16u},
\end{align}
we obtain
\begin{equation}
C_2(x,\theta)= C_2(\theta) x^2 +o(x^2)
\end{equation}
with
\begin{align}\label{C2}
&C_2(\theta)\equiv -\frac{\cos\theta}{4\sqrt{\pi}} \left(1-\frac{\pi^2}{\alpha^2}\right) \nonumber \\
& \times \int_0^{+\infty} \frac{du}{u} \left( \kern-0.3em \e{-u}I_{\nu(0)}(u)-\frac{u^{\nu(0)}}{2^{\nu(0)} \Gamma(1+\nu(0))} \right). 
\end{align}
Finally, we have obtained the development of the term $A_0$ up to order 2
\begin{align}
&T_0=\frac{4\theta\cos\theta}{\sqrt{\pi}(\pi-2\theta)}-\frac{2}{\pi}\sin\theta \, x +C(\theta) \, x^{2+2\nu(0)}\nonumber \\
& \qquad + C_2(\theta) \, x^2+ o(x^2).
\end{align}

\begin{figure}
\begin{minipage}{0.47\linewidth}
\centering
\includegraphics[width=120pt]{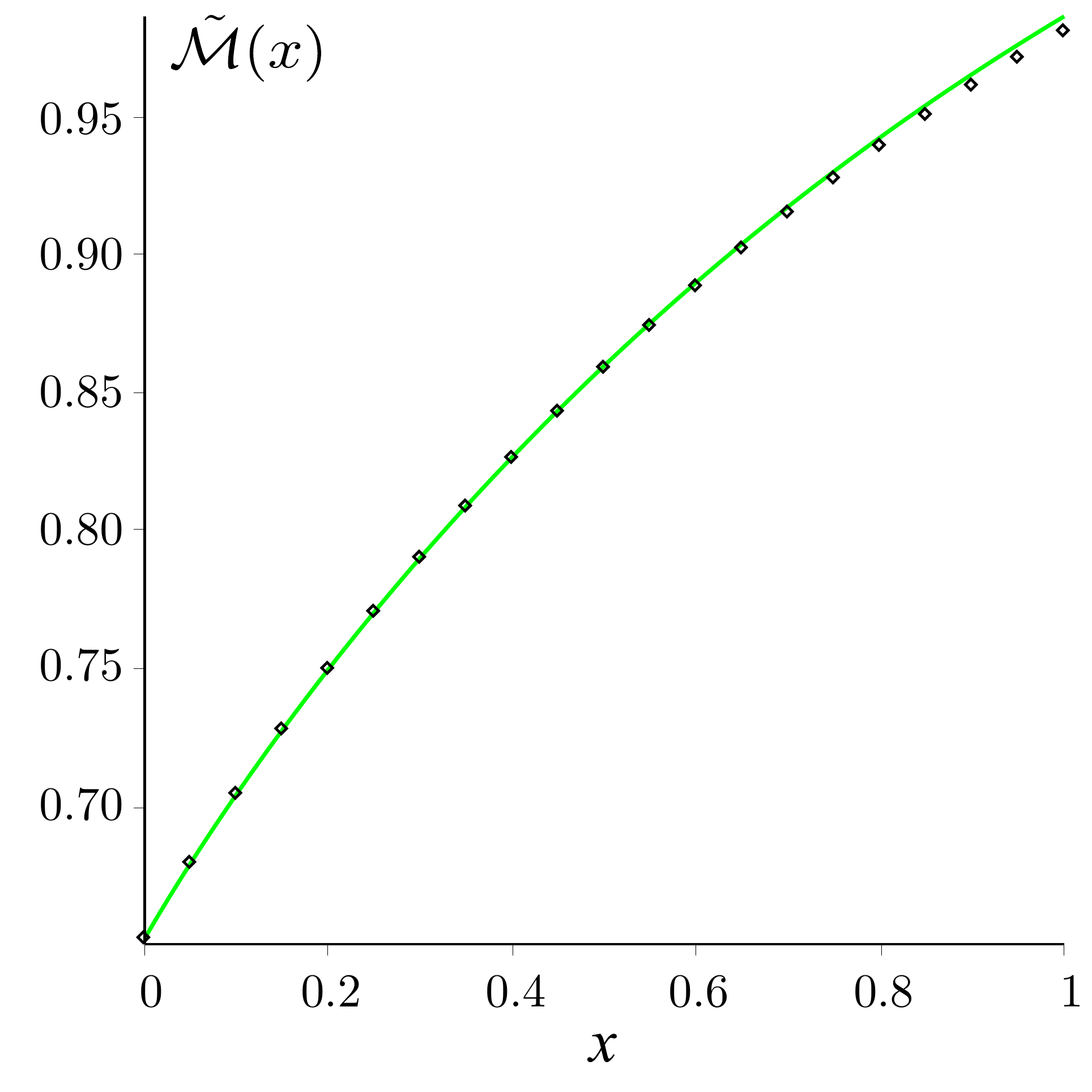}
\end{minipage}\hfill
\begin{minipage}{0.47\linewidth}
\centering
\includegraphics[width=120pt]{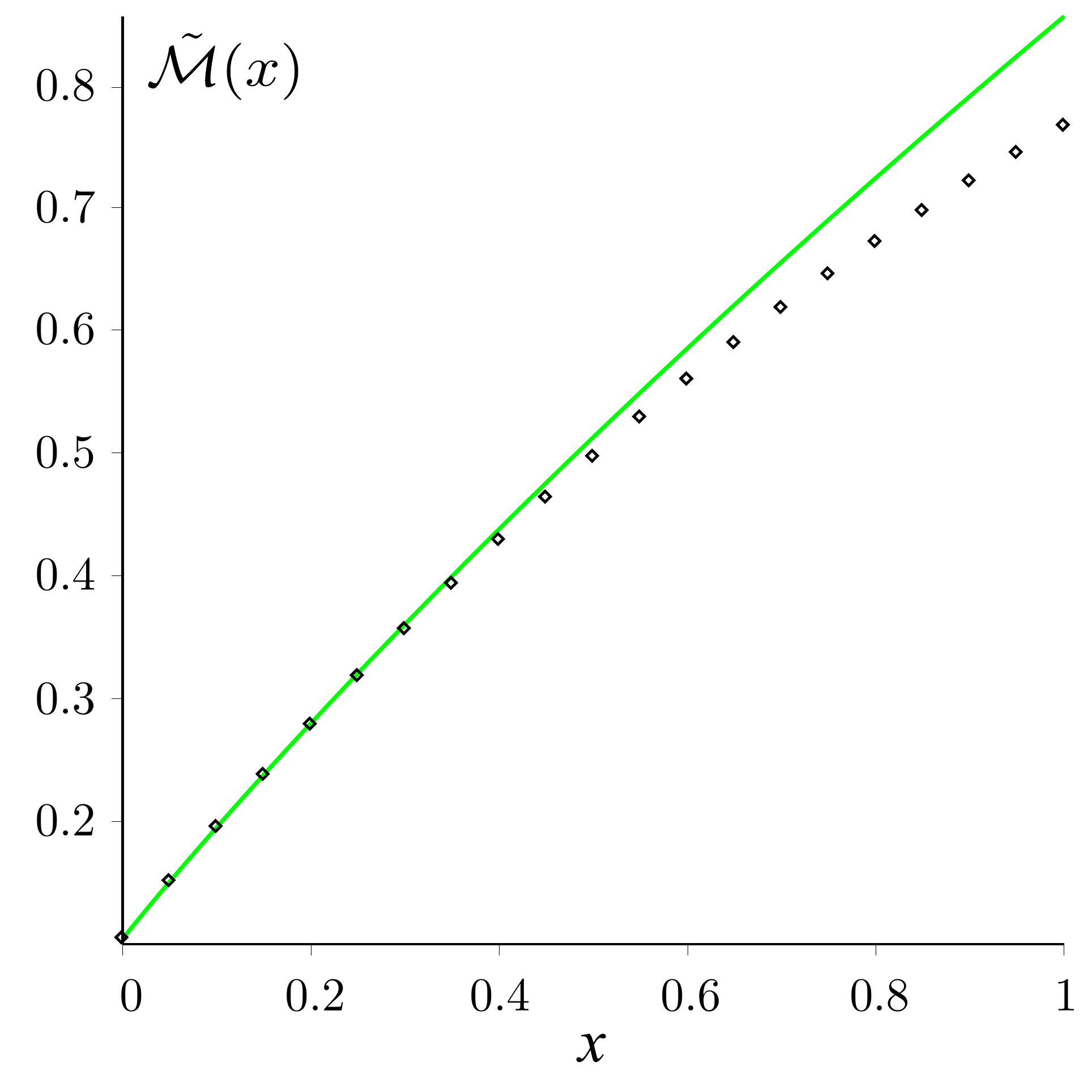}
\end{minipage}\hfill
\caption{Plot of the exact expression of the mean rescaled maximum and of the corresponding development given in Eq.~\eqref{Mnegdev}, for \mbox{$\theta=-\pi/5$} (left) and \mbox{$\theta=-4\pi/9$} (right). The range of validity of the development becomes smaller when $\theta$ approaches \mbox{$-\pi/2$}.}
\label{devM}
\end{figure}

Contrary to the determination of $T_0$, the calculation of the two remaining terms $T'_0$ and $T_1$ of Eq.~\eqref{defT} is straightforward. Indeed, the approach presented in \ref{thetapos} is valid for these terms as the indices of the involved Bessel functions (\mbox{$\nu(0)+1$} for $T'_0$ and \mbox{$\nu(m)$} with \mbox{$m\leqslant 1$} for $T_1$) are positive, and yields
\begin{align}
T'_0(\theta,x)=&\frac{4(1+\nu(0))\cos\theta}{\sqrt{\pi}} - \frac{2\sin\theta}{\pi} \, x \nonumber \\
& +\frac{\theta \cos\theta}{\sqrt{\pi}(\pi-2\theta)} \, x^2 +o(x^2) \\
T_1(\theta,x)=&-\frac{2\sqrt{\pi}\cos\theta}{\pi-2\theta} -\left(1-\frac{4}{\pi}\right) \sin\theta \, x \nonumber \\
& -\frac{\sqrt{\pi}\cos\theta}{2(\pi-2\theta)} \, x^2 +o(x^2). 
\end{align}
This gives the development of \mbox{$\tilde{M}(\theta,x)$} up to order 2
\begin{align}\label{Mnegdev}
&\tilde{M}(\theta,x)=\frac{2\sqrt{\pi}\cos\theta}{\pi-2\theta} -\sin\theta \, x + C(\theta) \, x^{2+\frac{2\theta}{\pi-2\theta}} \nonumber \\
& \qquad + \left( C_2(\theta)-\frac{\cos\theta}{2\sqrt{\pi}} \right) x^2 + o(x^2) 
\end{align}
where we replaced $\nu(0)$ and $\alpha$ with their values \mbox{$\theta/\alpha$} and \mbox{$\pi-2\theta$}.  The coefficients \mbox{$C(\theta)$} and \mbox{$C_2(\theta)$} are given by Eqs.\eqref{defC} and \eqref{C2}. As shown in Fig.~\ref{devM}, the range of validity of this development is large when $\theta$ is close to 0 and becomes smaller and smaller when $\theta$ approaches \mbox{$-\pi/2$}.

\subsubsection{Development of the mean perimeter}

\begin{figure}
\centering
\includegraphics[width=180pt]{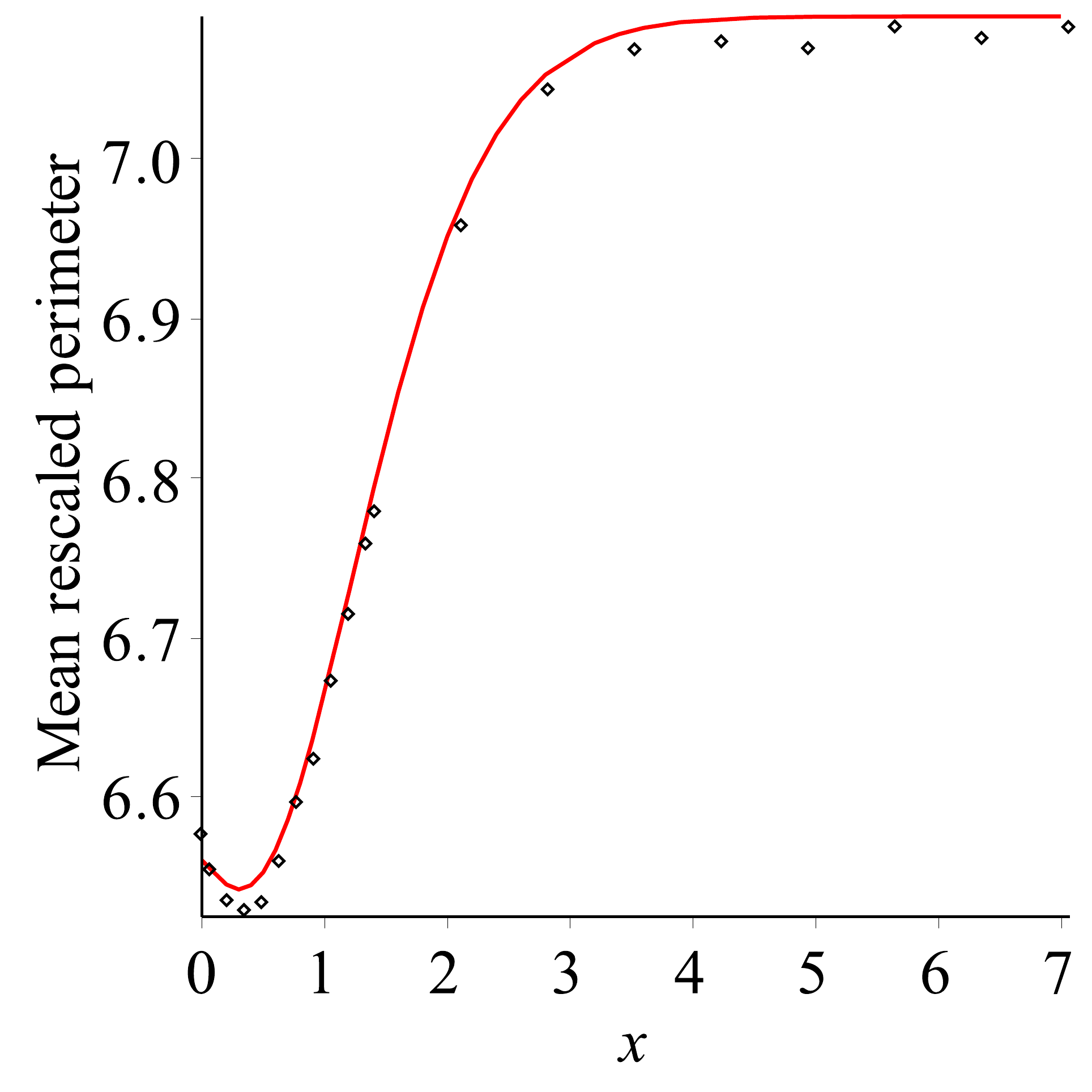}
\caption{Plot of the mean rescaled perimeter of the convex hull of a Brownian motion in the presence of a reflecting wall as a function of the rescaled initial distance $x$, compared to numerical simulations. The main feature of this observable is to display a minimum with respect to the rescaled initial distance to the wall. Some details on the numerical evaluation and simulations are given in Sections \ref{eval} and \ref{simus}.}
\label{MRP}
\end{figure}

We recall that
\begin{equation}
\tilde{L}(x)=2 \int_{-\pi/2}^{\pi/2} d\theta \; \tilde{\mathcal{M}}(\theta,x).
\end{equation}
The development of \mbox{$\tilde{\mathcal{M}}(\theta,x)$} for small $x$ is given in Eqs.~\eqref{Mposdev} and \eqref{Mnegdev} (respectively for \mbox{$\theta>0$} and \mbox{$\theta<0$}). By integration over $\theta$, the linear term disappears so the two-term development of the mean perimeter will stem from the following expression
\begin{equation}\label{devL1}
\tilde{L}(x)=2 \sqrt{\pi} \mathrm{Si}(\pi)+2\int_{-\pi/2}^0 d\theta \; C(\theta) \; x^{2+2\frac{\theta}{\pi-2\theta}}+ O(x^2).
\end{equation}
The integral over $\theta$ is dominated by the neighborhood of \mbox{$-\pi/2$} for small $x$ since the power of $x$ is a decreasing function of $\theta$. However, the first and third terms of \mbox{$C(\theta)$} involved in Eq.~\eqref{defC}, which have a finite limit in \mbox{$\theta=-\pi/2$}, disappear as \mbox{$\cos\theta$} goes to zero when $\theta$ goes to \mbox{$-\pi/2$}. In this limit, \mbox{$C(\theta)$} is then dominated by its second term
\begin{align}
&C(\theta) \equi{\theta \to -\frac{\pi}{2}}  - \frac{\cos\theta}{2^{4\nu(0)+1} \sqrt{\pi} \; \Gamma(1+\nu(0))} \nonumber \\ 
& \times \int_1^{\frac{1}{\cos^2\theta}} \kern-0.5em dz \sqrt{\frac{z}{z-1}} z^{\nu(0)} \cos\left(\frac{\pi}{\alpha}\arccos\left(-\frac{\sqrt{z-1}}{\sqrt{z}}\right) \right) 
\end{align}
with \mbox{$\alpha=\pi-2\theta$}. Setting \mbox{$\epsilon=\theta+\pi/2$} and introducing the new variable \mbox{$v=z \epsilon^2$}, we can write
\begin{align}\label{Ceps1}
C(\epsilon) \equi{\epsilon\to 0} &-\frac{\epsilon^{-1-2\nu(0)}}{2^{4\nu(0)+1}\sqrt{\pi} \; \Gamma(1+\nu(0))}  \int_{\epsilon^2}^1 dv \sqrt{\frac{v}{v-\epsilon^2}} v^{\nu(0)}\nonumber \\
& \times \cos\left(\frac{\pi}{2\pi-2\epsilon} \arccos\left(-\frac{\sqrt{z-1}}{\sqrt{z}}\right) \right).
\end{align}
In the limit \mbox{$\epsilon\to 0$}, 
\begin{equation}
\cos\left(\frac{\pi}{2\pi-2\epsilon} \arccos\left(-\frac{\sqrt{z-1}}{\sqrt{z}}\right) \right) \sim \frac{\epsilon}{2} \left(\frac{1}{\sqrt{v}}-1 \right)
\end{equation}
and Eq.~\eqref{Ceps1} becomes
\begin{equation}
C(\epsilon) \equi{\epsilon\to 0}  \frac{\sqrt{\epsilon}}{2\sqrt{\pi} \; \Gamma(3/4)} \int_{\epsilon^2}^1 dv \left(\frac{1}{\sqrt{v}}-1 \right) v^{-1/4}
\end{equation}
and finally
\begin{equation}
C(\epsilon) \sim -\frac{4\sqrt{\epsilon}}{3\sqrt{\pi} \; \Gamma(3/4)}.
\end{equation}
We can then rewrite
\begin{align}
&\int_{-\pi/2}^0 d\theta \; C(\theta) \; x^{2+2\frac{\theta}{\pi-2\theta}} \sim \nonumber \\
& -\frac{4}{3\sqrt{\pi} \; \Gamma(3/4)}\int_{-\pi/2}^0 d\theta \sqrt{\theta+\frac{\pi}{2}} \exp\left[\left(2+\frac{2\theta}{\pi-2\theta}\right) \ln x\right] .
\end{align}
For small $x$, \mbox{$|\ln x|$} is large. Using Laplace's method, we can rewrite this integral as follows, after the change in variables \mbox{$\epsilon=\theta+\pi/2$}
\begin{align}
\int_0^{+\infty} \kern-0.5em d\epsilon \sqrt{\epsilon} \exp\left(-\frac{\epsilon}{2\pi} \ln \frac{1}{x}\right) \sim  \frac{\sqrt{2}\pi^2}{\left(\ln \frac{1}{x}\right)^{3/2}}
\end{align}
so the development of the mean rescaled perimeter at small distance $x$ is
\begin{equation}
\tilde{L}(x)-2 \sqrt{\pi} \mathrm{Si}(\pi) \equi{x\ll1} -\frac{8 \sqrt{2 \pi^3}}{3\Gamma(3/4)} \frac{x^{3/2}}{\left(\ln \frac{1}{x}\right)^{3/2}},
\end{equation}
which turns out to be non-analytical. Moreover, the mean rescaled perimeter has a value in \mbox{$x=0$} that is lower than its limits when $x$ is large (corresponding to the case without any confinement), as mentioned previously, and is a decreasing function at small $x$. This implies that the mean rescaled perimeter displays a minimum with respect to the initial distance to the wall, as confirmed by the numerical evaluation of its exact expression and by the numerical simulations (see Fig.~\ref{MRP}). We point out that this small $x$ development has an extremely small range of validity. Indeed, as we previously noticed, the closer $\theta$ to \mbox{$-\pi/2$}, the smaller the range of validity of the development of \mbox{$\tilde{\mathcal{M}}(x)$}. Since the development of \mbox{$\tilde{L}(x)$} is dominated by the contributions of the mean maximum for directions $\theta$ close to \mbox{$-\pi/2$}, the range of validity of the mean perimeter is substantially affected. Nevertheless, the sign of the first correction is sufficient to conclude on the non-monotonicity of the mean rescaled perimeter.

\subsection{Numerical evaluation of the mean perimeter}\label{eval}

In practice, the numerical evaluation of the expression of the mean perimeter given in Eq.~\eqref{LBessel} turns out to be tricky. The mean maximum in a given direction $\theta$ itself, which is the integrand of the integral over $\theta$ involved in Eq.~\eqref{LBessel}, cannot be computed straightforwardly since it already involves an infinite integral of an infinite sum of special functions. To bypass this difficulty, we use a lighter alternative expression of the survival probability of a Brownian particle in an infinite absorbing wedge derived in \cite{Chupeau:2015a}. For acute wedges of top angle \mbox{$\alpha<\pi$}, the survival probability is given by
\begin{align}\label{acute}
& S(y,\varphi_0) = \erf{\sqrt{2y}\sin\left(\varphi_0 \right)}\nonumber \\
& +\sum\limits_{j=1}^{k} (-1)^j \left[\erf{\sqrt{2y}\sin\left(j\alpha+\varphi_0\right)} \right.\nonumber \\
&\qquad  \left. -\erf{\sqrt{2y}\sin\left(j\alpha-\varphi_0\right)} \right]  \nonumber \\
& +(-1)^{k+1} \left[ \erf{\sqrt{2y}\sin\left(\min\left((k+1)\alpha+\varphi_0,\frac{\pi}{2}\right)\right)} \right.\nonumber \\
&\qquad  \left. -\erf{\sqrt{2y}\sin\left(\min\left((k+1)\alpha-\varphi_0,\frac{\pi}{2}\right)\right)} \right]  \nonumber \\
&  \kern-1em +\left(\frac{2}{\pi}\right)^{3/2} \sqrt{y} \; \frac{e^{-y}}{2} \int_0^{+\infty} dv \; e^{-y\cosh v} \sinh\frac{v}{2} \times \nonumber \\
&\kern-1em  \left[ \arctan\left( \frac{\sin\left(\frac{\pi}{\alpha}\left(\varphi_0+\frac{\pi}{2}\right)\right)}{\sinh\left(\frac{\pi v}{2\alpha}\right)}\right) +\arctan\left( \frac{\sin\left(\frac{\pi}{\alpha}\left(\varphi_0-\frac{\pi}{2}\right)\right)}{\sinh\left(\frac{\pi v}{2\alpha}\right)}\right)  \right] \nonumber \\
\end{align}
and for obtuse wedges of top angle \mbox{$\alpha>\pi$}, by
\begin{align}\label{obtuse}
&S(y,\varphi_0) = \mathrm{erf}\left(\sqrt{2y}\sin\left(\min\left(\varphi_0,\frac{\pi}{2}\right) \right)\right) \nonumber \\
&\kern-1em+\left(\frac{2}{\pi}\right)^{3/2} \sqrt{y} \; \frac{e^{-y}}{2} \int_0^{+\infty} dv \; e^{-y\cosh v} \sinh\frac{v}{2}  \times \nonumber \\
& \kern-1em \left[ \arctan\kern-0.2em \left( \frac{\sin\left(\frac{\pi}{\alpha}\left(\varphi_0+\frac{\pi}{2}\right)\right)}{\sinh\left(\frac{\pi v}{2\alpha}\right)}\right) \kern-0.3em+\arctan\kern-0.2em\left( \frac{\sin\left(\frac{\pi}{\alpha}\left(\varphi_0-\frac{\pi}{2}\right)\right)}{\sinh\left(\frac{\pi v}{2\alpha}\right)}\right) \kern-0.2em \right], \nonumber \\
\end{align}
with \mbox{$k=\lfloor\pi/(2\alpha) -1/2\rfloor$} and \mbox{$y=r_0^2/(8Dt)$}. We recall that the parameters $\alpha$, $r_0$ and $\varphi_0$ can be written in terms of the natural rescaled variables and parameter of the initial convex hull problem $u$, $x$ and $\theta$
\begin{align}
& \alpha=\pi-2\theta \nonumber \\
& \frac{r_0}{\sqrt{Dt}}=\frac{1}{\cos\theta} \sqrt{x^2+2 x u \sin\theta+u^2} \nonumber \\
& \varphi_0=\arccos\left( \frac{x+u \sin\theta}{\sqrt{u^2+2xu\sin\theta+x^2}}\right). \nonumber
\end{align}
The expressions of the survival probability \eqref{acute} and \eqref{obtuse} have the double advantage to involve a \textit{finite} sum of \textit{elementary} functions, as opposed to the infinite sum of Bessel functions given by Eq.~\eqref{survie}. This step is determining as we now manage to evaluate the double integral over the rescaled distance $u$ and the direction $\theta$ of the survival probability involved in Eq.~\eqref{defperim}. Note that in practice we need to truncate the integrals over $u$ and over $v$, the latter appearing in the last term of the survival probability, to carry out the numerical evaluation.

\subsection{Numerical simulations}\label{simus}

We compare our theoretical expression of the mean perimeter of the convex hull with numerical simulations. We generated Gaussian random walks of $10^5$ steps with a constant time step \mbox{$\Delta \tau=10^{-3}$} when the walker is farther than a distance \mbox{$d\simeq 0.2$} from the reflecting wall. When the walker approaches the wall, the time step is adapted, taken quadratic in the distance $d$ to the wall \mbox{$\Delta \tau = (0.1 \, d+\lambda)^2$} with \mbox{$\lambda=0.01$}. The cutoff $\lambda$ must not be too small to prevent the computation time from diverging. When the walker crosses the reflecting wall, we reflect it on the wall following the Snell-Descartes law of reflection. Once the trajectory has been constructed, the convex hull is then built using the Graham scan algorithm (see \cite{Graham:1972} or \cite{randon:these}), its perimeter calculated and averaged over $10^5$ realizations. Agreement is found with our analytical prediction (see Fig.~\ref{MRP}). Note that one has to make a compromise between the computation time and the precision of the generated trajectory (the steps must be as short as possible to approach at best a Brownian trajectory, especially near the wall).

\section{Physical mechanisms underlying the minimum of the mean perimeter of the convex hull}

As seen above, the effect of the confinement on the mean perimeter of the convex hull of the trajectory is non trivial, as it produces a non-monotonicity with respect to the initial distance to the reflecting wall. In this section, we give some keys to understand the underlying physical mechanisms. 

\begin{figure}
\centering
\includegraphics[width=180pt]{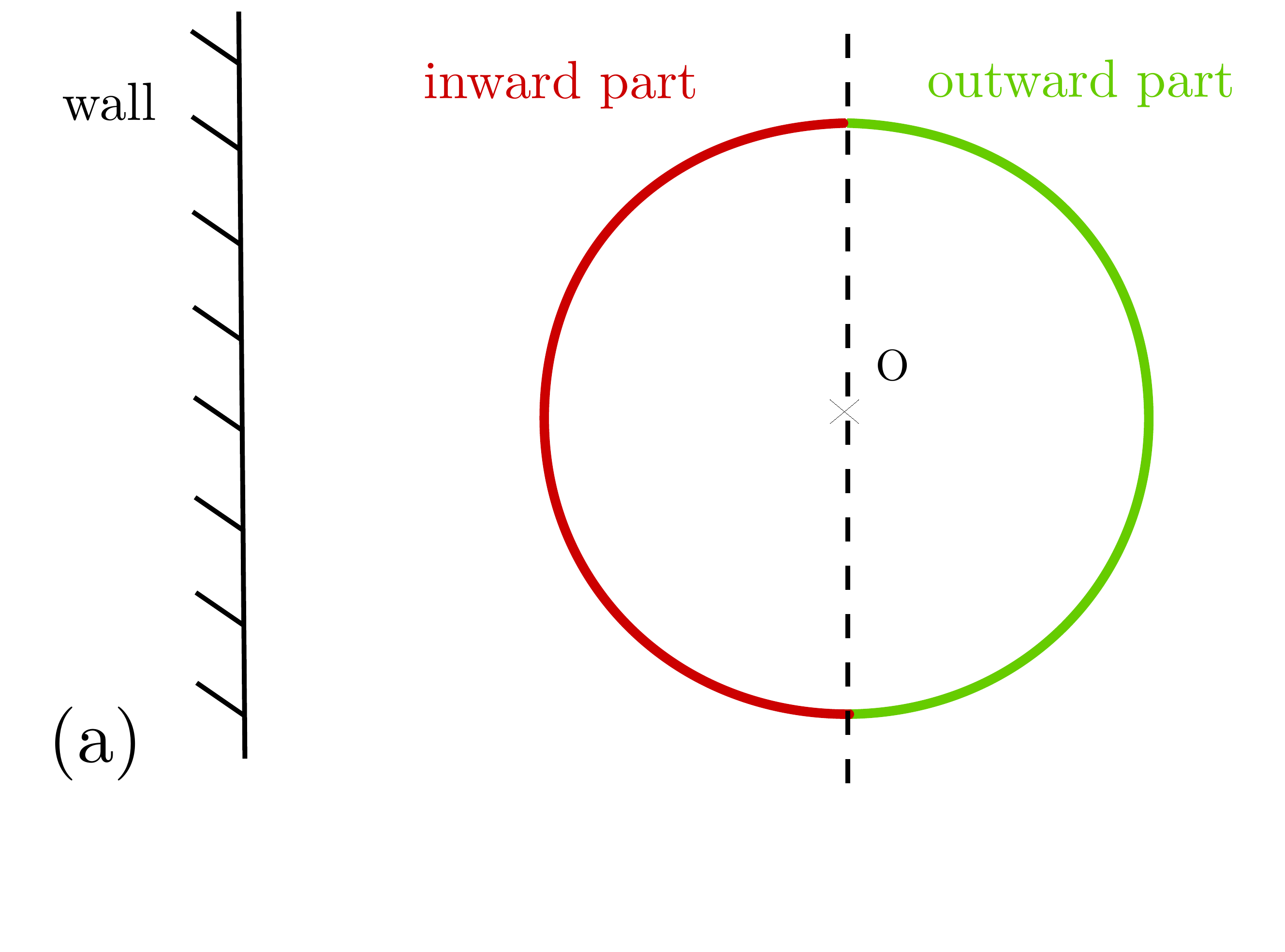}
\begin{minipage}{0.45\linewidth}
\includegraphics[width=120pt]{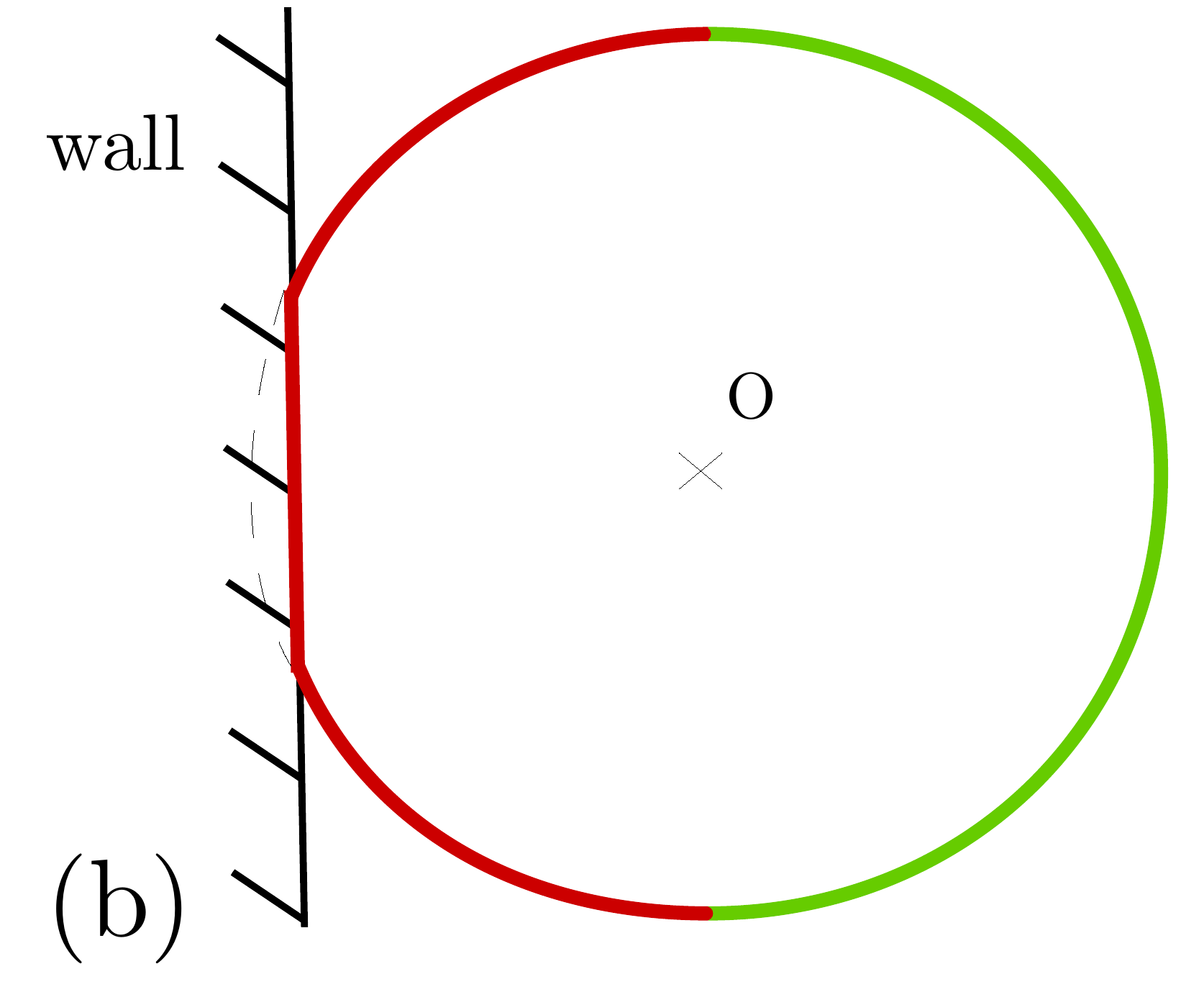}
\end{minipage}\hfill
\begin{minipage}{0.45\linewidth}
\includegraphics[width=100pt]{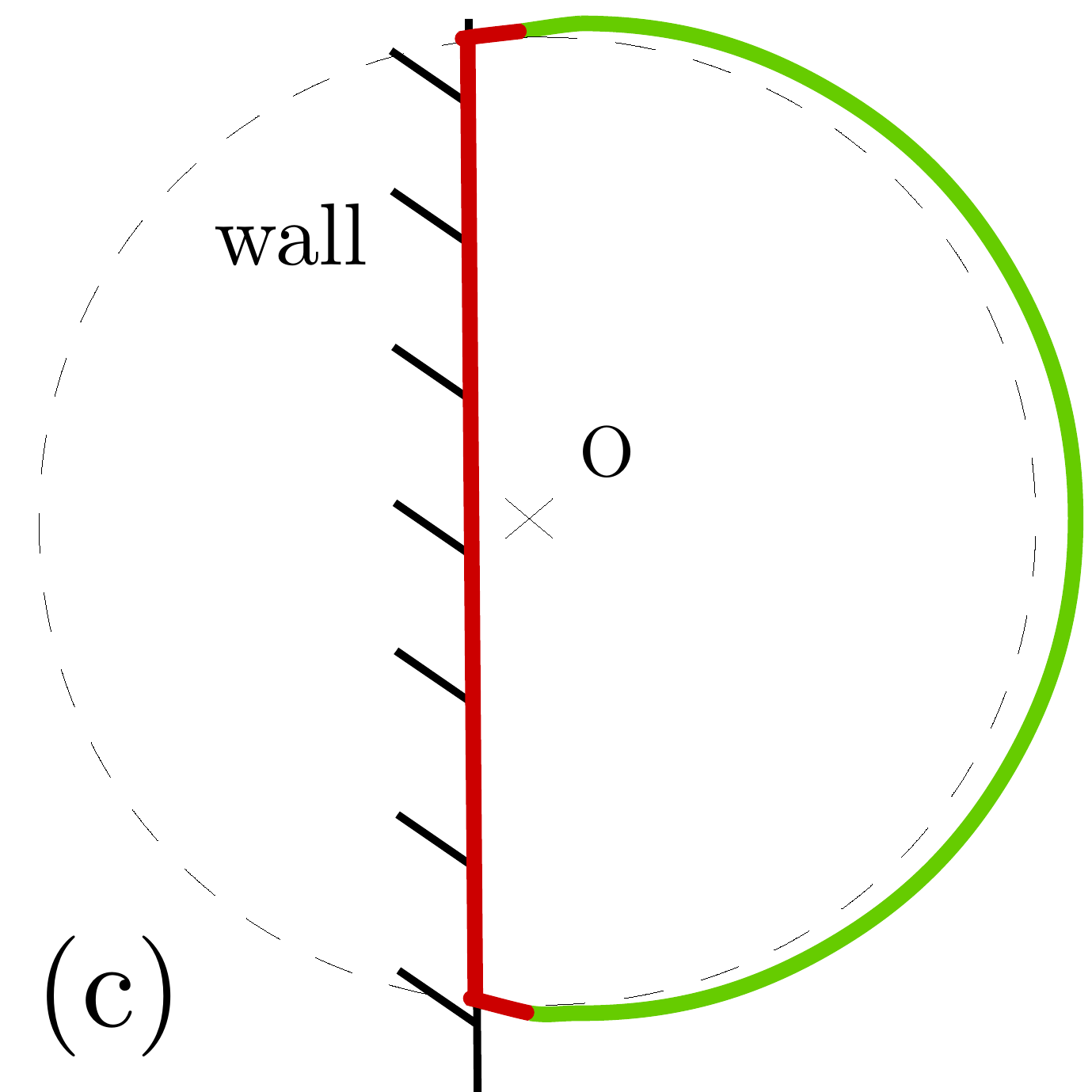}
\end{minipage}
\caption{Schematic representation of the convex hull, divided into two parts, an inward part in red and an outward part in light green. The wall has two effects, an effect of reduction of the accessible space that acts on the inward part of the convex hull, and an effect of effective repulsion that impacts on the outward part of the convex hull. (a) The trajectories feel no effect of the wall, the two parts of the convex hull have the same length. (b) The effect of reduction of the accessible cuts the inward part of the convex hull. (c) The length of the inward part is lower and lower, but the effect of effective repulsion swells the outward part.}
\label{2effects}
\end{figure}

The minimum of the mean perimeter can be interpreted as the result of the competition between two antagonistic effects of the wall on the trajectories: reduction of the accessible space and effective repulsion of the trajectories. To understand how these two effects are at play, let us split the convex hull into two parts, delimited by a line parallel to the wall passing through the starting point. It defines an inward part (between the wall and the line, in red in Fig.~\ref{2effects}) and an outward part (in light green in Fig.~\ref{2effects}). Three following cases emerge. 

First, if the starting point is far from the wall (\mbox{$d \gg \sqrt{Dt}$} with $t$ the observation time), at time $t$ the trajectories have not touched the wall. As in the non-confined geometry, the two parts of the convex hull, schematically represented as a circle of radius $\sqrt{Dt}$ by symmetry, have the same length \mbox{$2\sqrt{\pi Dt}$} (see Fig.~\ref{2effects} (a)). 

Then, if the starting point is such that \mbox{$d\sim \sqrt{Dt}$}, at time $t$ the trajectories start to feel the presence of the wall, which blocks the trajectories. This is a first effect of the wall, a \textit{reduction of the accessible space}. Consequently, the inward part of the convex hull is cut (see Fig.~\ref{2effects} (b)), and its length is lower than without confinement. The closer to the wall the starting point, the shorter this part of the convex hull. 

Finally, if the starting point is close to the wall (\mbox{$d \ll \sqrt{Dt}$}), the effect of reduction of the accessible space still exists, but another more subtle effect of the wall appears. Indeed, as the wall blocks the trajectories in one direction, it also pushes them in the opposite direction. Therefore, the outward part of the convex hull gets more rounded (see Fig.~\ref{2effects} (c)), so its length is higher than in the non-confined geometry. This is the second effect of the wall, an \textit{effective repulsion} of the trajectories, which is, as the first effect, more substantial when the initial distance is small. The combination of these two antagonistic effects yields the minimum of the mean perimeter of the convex hull.

\begin{figure}
\centering
\includegraphics[width=150pt]{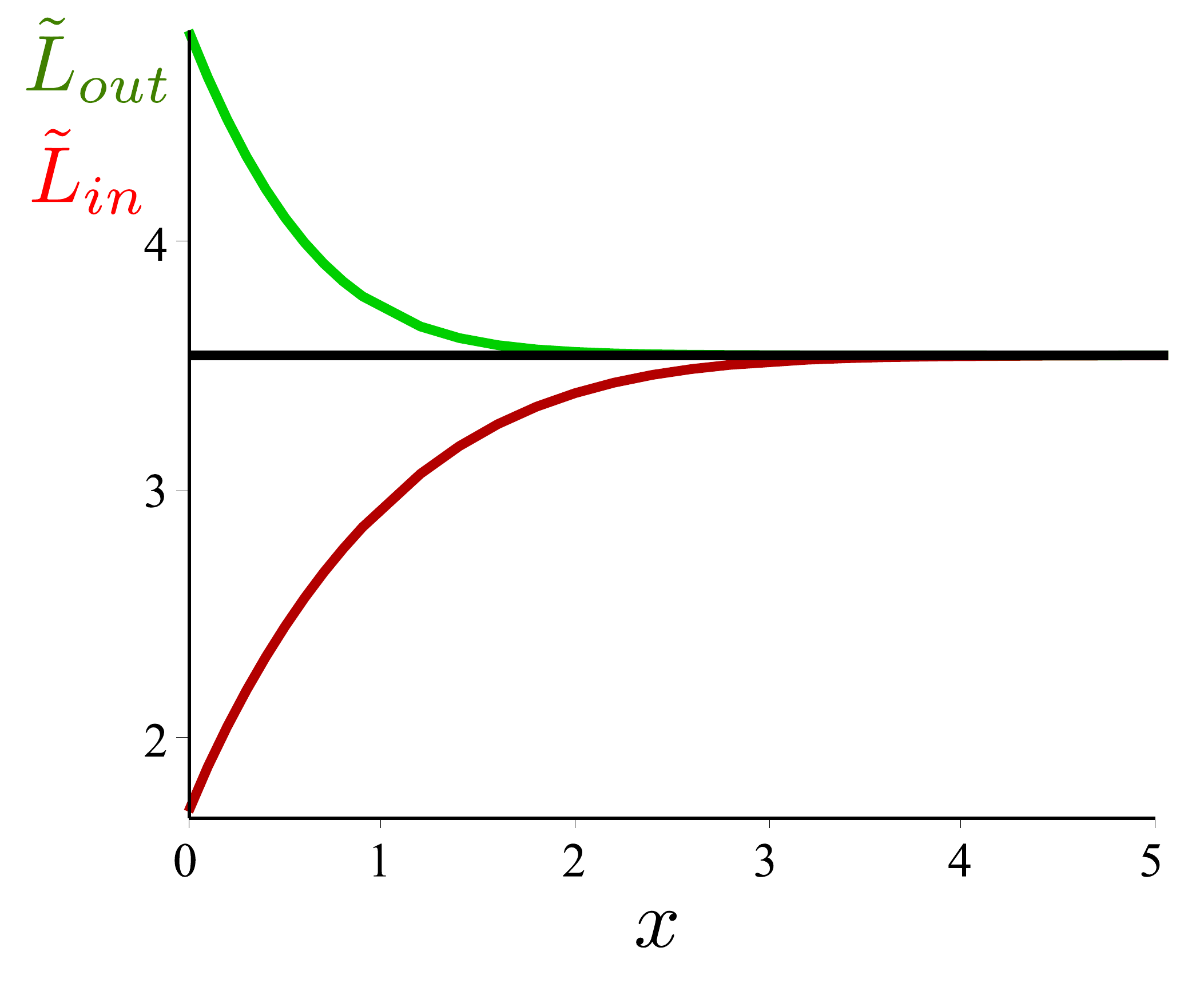}
\caption{Rescaled length of the inward (in red) and outward (green) parts of the convex hull with respect to the rescaled initial distance to the wall, obtained from Eqs.~\eqref{defportions}. The black horizontal line represents the length of these two parts in absence of confinement (\mbox{$2\sqrt{\pi}$}).}
\label{quant}
\end{figure}

Quantitatively, we check that the lengths of these two parts of the convex hull are respectively an increasing function of the initial distance for the inward part, and a decreasing function of the initial distance for the outward part (see Fig.~\ref{quant}). These two quantities are defined as follows
\begin{align}\label{defportions}
& \tilde{L}_{in}(x)  = 2\int_{-\pi/2}^0 d\theta \; \tilde{\mathcal{M}}(\theta,x) \\
& \tilde{L}_{out}(x) =2 \int_{0}^{\pi/2} d\theta \; \tilde{\mathcal{M}}(\theta,x)
\end{align}
where \mbox{$\tilde{\mathcal{M}}(\theta,x)=\langle\mathcal{M}^{(d)}(\theta,t)\rangle/\sqrt{Dt}$} is obtained from Eqs.~\eqref{defmax}, \eqref{acute} and \eqref{obtuse}. This graph also shows that the wall has a larger range of influence on the inward part of the convex hull (approximately \mbox{$d<3\sqrt{Dt}$} or \mbox{$x<3$}) than on the outward part (\mbox{$d<2\sqrt{Dt}$} or \mbox{$x<2$}), as expected from the previous qualitative discussion.

\begin{figure}
\centering
\includegraphics[width=170pt]{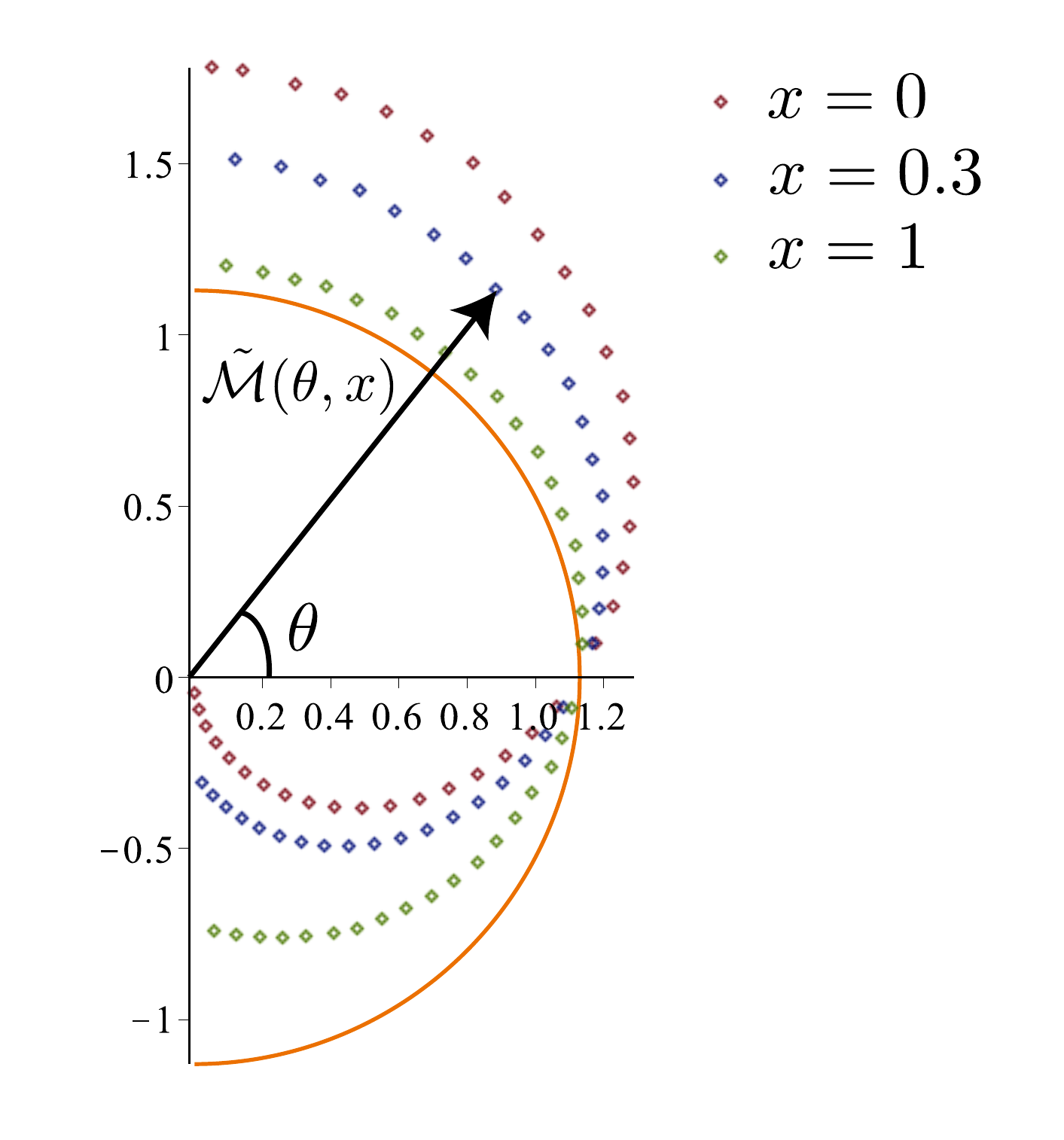}
\caption{Plot of the mean rescaled maximum in the direction $\theta$ parametrized by the angle $\theta$, obtained from Eqs.~\eqref{defmax}, \eqref{acute} and \eqref{obtuse}. The dots represent the mean rescaled maximum for 3 values of the initial rescaled distance, and the plain circle gives the mean rescaled maximum for a non-confined walk (which is \mbox{$2/\sqrt{\pi}\simeq 1.13$}).}
\label{nuage}
\end{figure}

It is interesting to notice that the qualitative arguments given above on the two portions of the convex hull can also apply to the mean maximum of the trajectory. Indeed, as presented in Fig.~\ref{nuage}, the mean maximum in the direction $\theta$ is higher than the non-confined value in the outward directions \mbox{$\theta>0$}, and lower than this value in the inward directions \mbox{$\theta<0$}, for any $\theta$. For the short initial distances, the increase of the mean maximum in the outward directions is substantial. The effective repulsion, which is a less obvious effect of the wall on the trajectories, is then important in the same way as the reduction of accessible space. Nevertheless, as mentioned before, the effective repulsion remains limited to shorter distances than the reduction of accessible space. Indeed, note that for \mbox{$x=1$}, the reduction of accessible space impacts on the directions \mbox{$\theta<0$} whereas the effective repulsion is quasi invisible on the directions \mbox{$\theta>0$} (see Fig.~\ref{nuage}).

\section{Mean extension of the visited portion of the reflecting wall}

\begin{figure}[h]
\centering
\includegraphics[width=140pt]{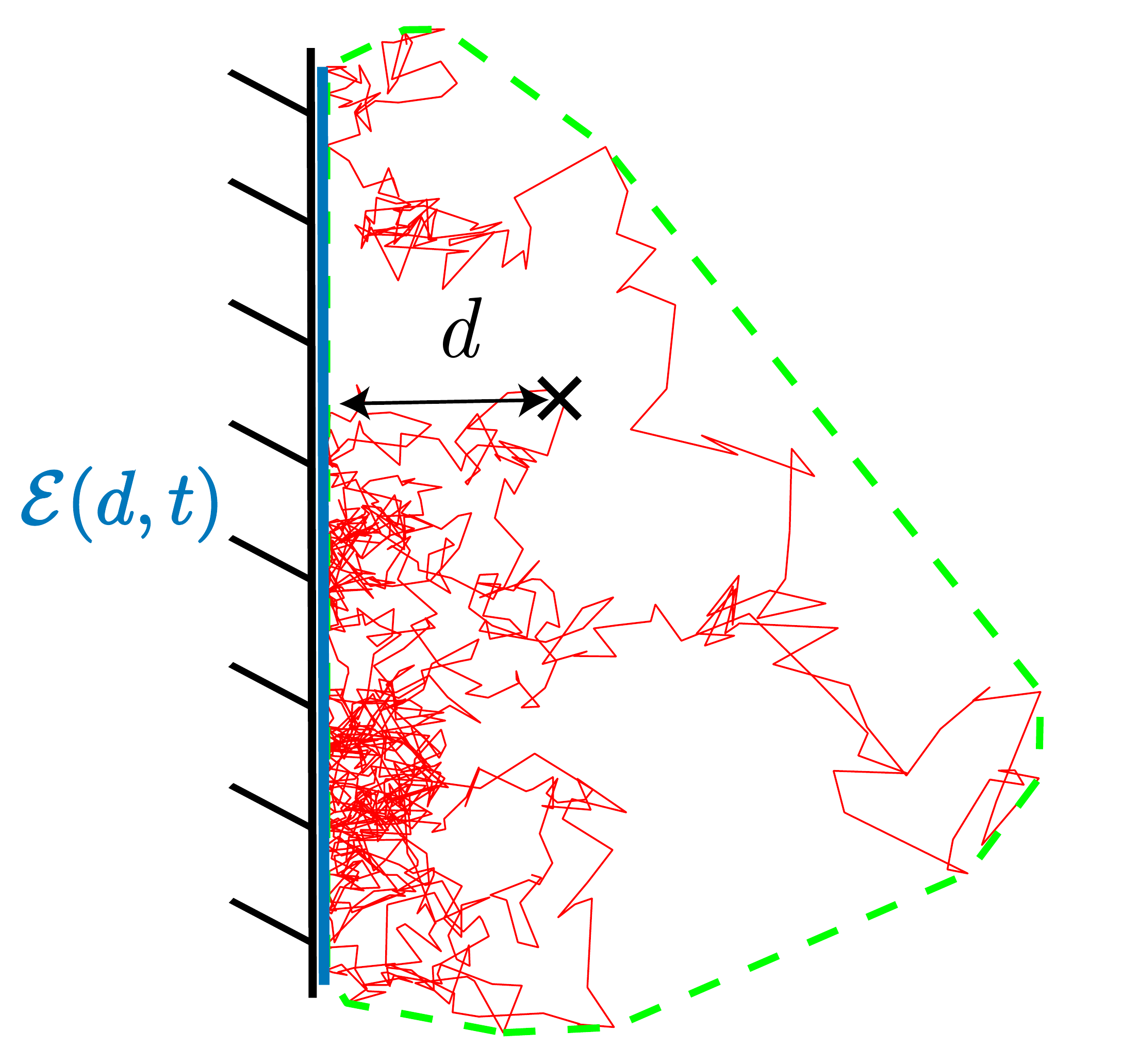}
\caption{Definition of the extension of the convex hull on the wall \mbox{$\mathcal{E}(d,t)$} at time $t$ for a Brownian motion starting at a distance $d$ from the wall, represented by the plain blue segment along the wall.}
\label{defext}
\end{figure}

A further quantification of the convex hull is obtained by focusing on the length of the part of the convex hull that is along the reflecting wall. It is defined as the maximal distance between two points where the Brownian motion has touched the wall (see Fig.~\ref{defext}). In this section, we determine the mean rescaled extension on the wall at time $t$ \mbox{$\tilde{\mathcal{E}}(x)\equiv \langle\mathcal{E}(d,t)\rangle /\sqrt{Dt}$} as a function of the initial rescaled distance to the wall $x$

\subsection{Particular case of a Brownian motion starting from the wall}

\begin{figure}
\centering
\includegraphics[width=170pt]{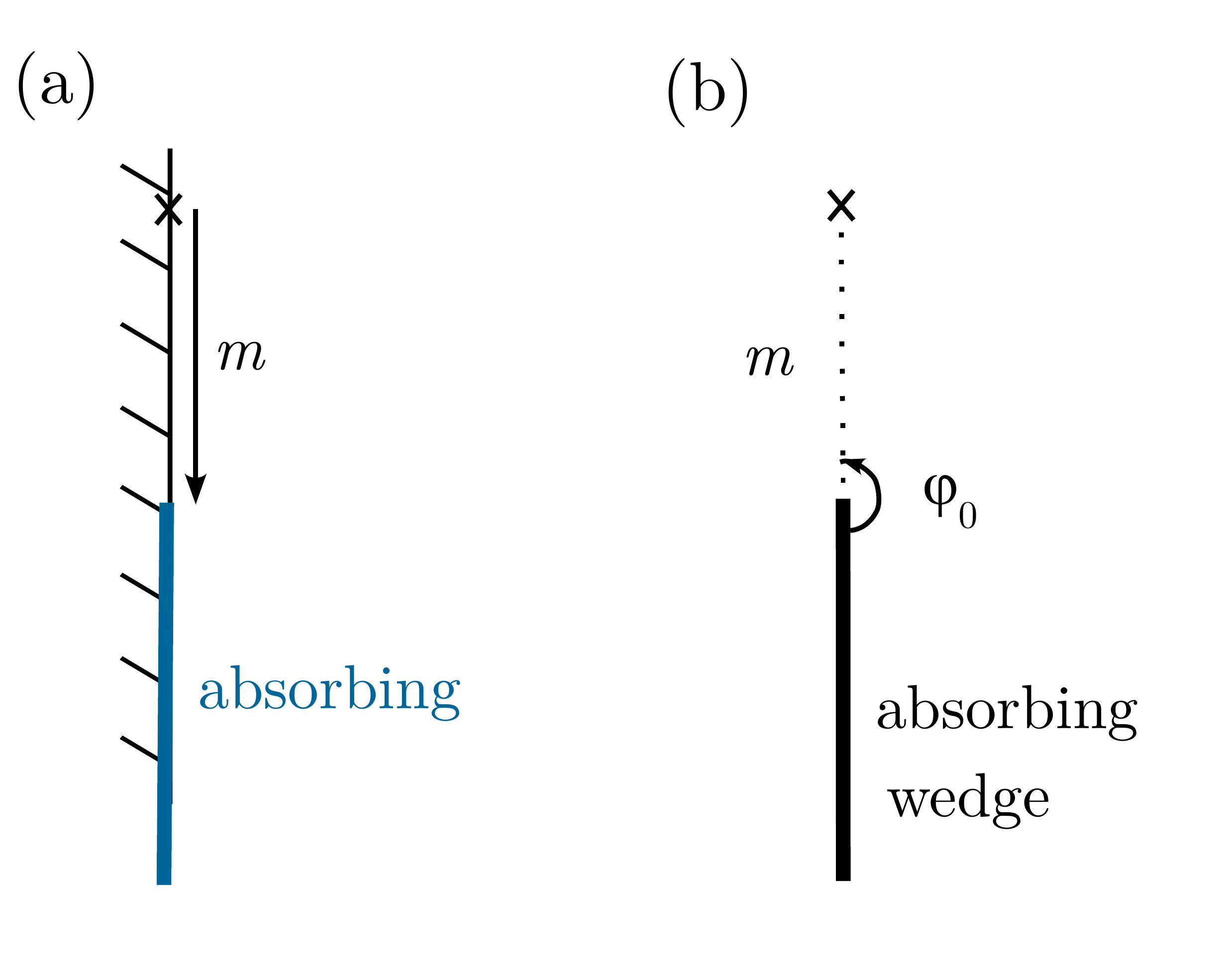}
\caption{(a) The probability that the rightward semi-extension \mbox{$\mathcal{E}_{\mathrm{half}}$} is smaller than $m$ is the survival probability in the presence of the semi-infinite absorbing thick blue line. (b) This probability is exactly the survival probability in an infinite absorbing wedge of top angle \mbox{$2\pi$} for a starting point located at a distance $m$ from the apex of the wedge with an angle \mbox{$\varphi_0=\pi$}.}
\label{wedge_ext}
\end{figure}
We first consider the particular case where the Brownian walker starts from the wall and determine the mean rescaled extension \mbox{$\tilde{\mathcal{E}}(0)$}. The probability that the semi-extension on the wall \mbox{$\mathcal{E}_{\mathrm{half}}$} is smaller than $m$ is exactly the probability not to have touched the absorbing semi-infinite line along the wall starting at a distance $m$ from the initial point (see Fig.~\ref{wedge_ext})
\begin{equation}
F(\mathcal{E}_{\mathrm{half}}<m)=S(m,\pi).
\end{equation}
 This probability \mbox{$S(m,\pi)$} is the survival probability in an absorbing wedge of top angle \mbox{$2\pi$} with a starting point at distance $m$ of the apex and an angle \mbox{$\varphi_0=\pi$} (see Fig.~\ref{wedge_ext}). Introducing as previously the rescaled distance \mbox{$u=m/\sqrt{Dt}$}, this survival probability is given by \cite{Chupeau:2015a}
\begin{align}
&S(u,\pi)=\erf{\frac{u}{2}} +\frac{u}{\pi^{3/2}} \exp\left( -\frac{u^2}{8} \right) \times\nonumber  \\
&\int_0^{+\infty} \kern-0.5em dv \exp\left( -\frac{u^2}{8} \cosh v\right) \sinh \frac{v}{2} \arctan \left( \frac{1}{\sqrt{2} \sinh \frac{v}{4}}\right).
\end{align}
Following the same lines as from Eq.~\eqref{defM2} to Eq.~\eqref{defmax}, the mean rescaled extension on the wall, which is twice the mean rescaled semi-extension, can be written as
\begin{equation}
\tilde{\mathcal{E}}(0)=2 \; \tilde{\mathcal{E}}_{\mathrm{half}}(0) = 2 \int_0^{+\infty} du \; (1-S(u,\pi)).
\end{equation}
This integral turns out to be doable analytically and leads to 
\begin{equation}\label{ext0}
\tilde{\mathcal{E}}(0) =\frac{2}{\sqrt{\pi}}\simeq 1.128.
\end{equation}
Note that this value is twice as small as the mean span of the trajectory in the direction parallel to the wall \mbox{$S(0)$} (see Eq.~\eqref{span0}).

\subsection{General case}

\begin{figure}[h]
\centering
\includegraphics[width=160pt]{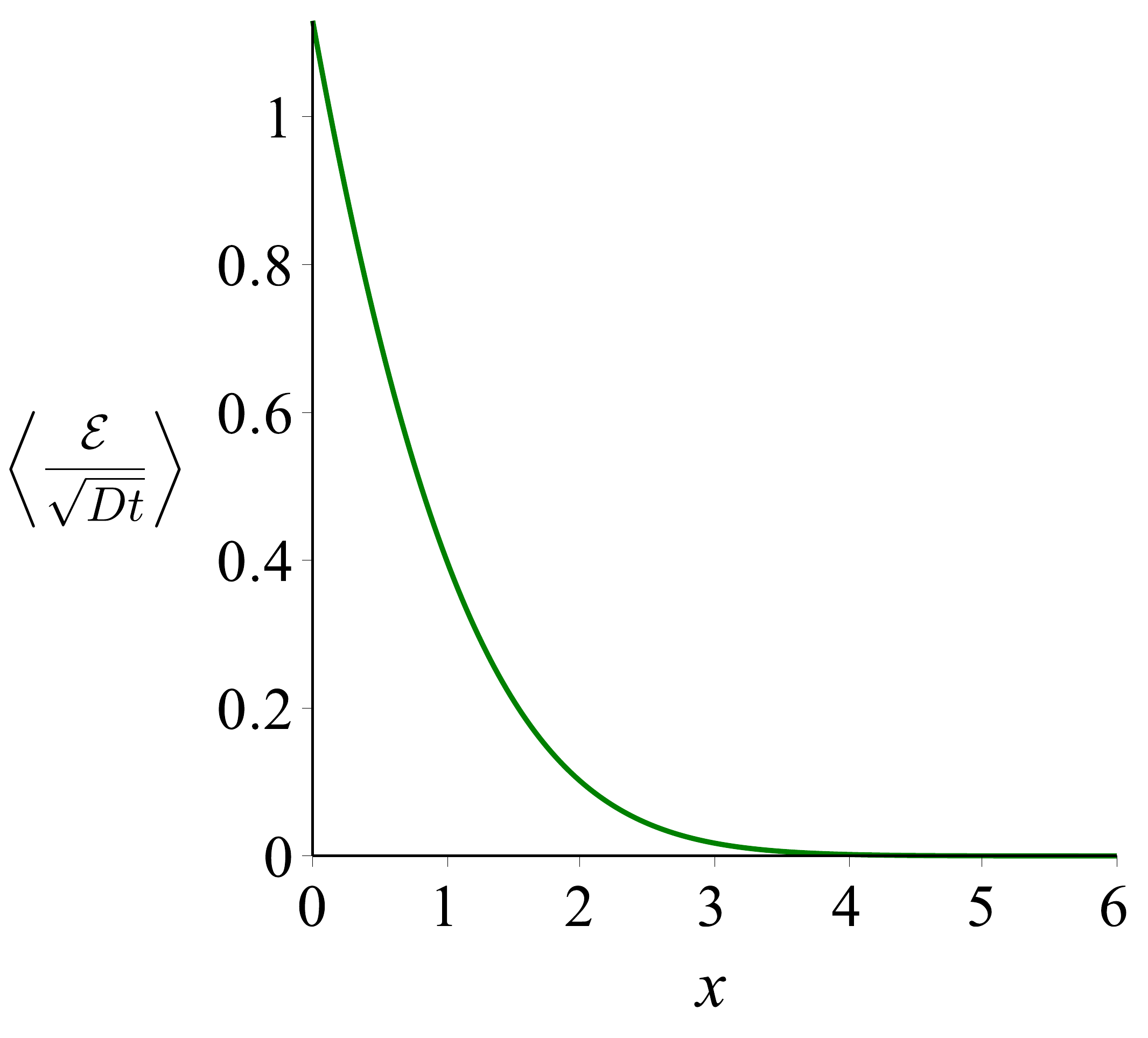}
\caption{Plot of the analytical expression \eqref{ext} of the mean extension at time $t$ of a planar Brownian motion on the reflecting wall rescaled by $\sqrt{Dt}$ as a function of the rescaled initial distance \mbox{$x=d/\sqrt{Dt}$}.}
\label{extension_courbe}
\end{figure}

We now determine the mean rescaled extension on the wall at time $t$ for the general case of a Brownian motion starting at a distance $d$ from the wall. If the trajectory does not touch the wall up to time $t$, this extension is zero. Otherwise, if the Brownian walker first touches the wall at a time \mbox{$t'<t$}, the mean extension of such a trajectory then reduces to the calculation of \mbox{$\langle \mathcal{E}(0,t-t') \rangle$}. Therefore, the mean extension of all trajectories is
\begin{equation}\label{convolution}
\langle \mathcal{E}(d,t) \rangle=\int_0^t dt' F(d,t_{abs}=t') \langle \mathcal{E}(0,t-t') \rangle
\end{equation}
with \mbox{$F(d,t_{abs}=t')$} the first-passage probability density to the wall at time $t'$ starting at a distance $d$ from the wall \cite{Redner,Bray:2013}
\begin{equation}\label{FPT}
F(d,t_{abs}=t')=\frac{d}{\sqrt{4\pi D t'^{3/2}}} \exp\left( -\frac{d^2}{4Dt'} \right).
\end{equation}
Eq.~\eqref{convolution} involves a convolution, so it is convenient to take the Laplace transform of the mean extension
\begin{equation}
\langle \hat{\mathcal{E}}(d,s) \rangle = \hat{F}(d,s) \langle \hat{\mathcal{E}}(0,s) \rangle.
\end{equation}
From Eqs.~\eqref{ext0} and \eqref{FPT}, we have 
\begin{equation}
\hat{F}(d,s)=\exp\left( -d \sqrt{\frac{s}{D}} \right)
\end{equation}
and
\begin{equation}
\langle \hat{\mathcal{E}}(0,s) \rangle=\sqrt{\frac{D}{s^3}}.
\end{equation}
After inverse Laplace transform, we obtain the rescaled mean extension on the wall at time $t$ as a function of the initial rescaled distance \mbox{$x=d/\sqrt{Dt}$}
\begin{equation}\label{ext}
\tilde{\mathcal{E}}(x)= \frac{2}{\sqrt{\pi}} \exp\left( -\frac{x^2}{4} \right) -x \erfc{\frac{x}{2}}
\end{equation}
which is plotted on Fig.~\ref{extension_courbe}. As expected, this is a decreasing function of the initial distance to the wall.

\section{Conclusion}

In this article, we established a scaling expression of the mean perimeter of the convex hull of a Brownian motion at time $t$, rescaled by $\sqrt{Dt}$, starting at a rescaled distance $x$ from an infinite reflecting wall. By carrying out a thorough analysis of this mean rescaled perimeter, we demonstrated that it is a non-monotonic function of the rescaled initial distance $x$. It means that there exists an optimal initial distance to the wall that minimizes the mean perimeter of the convex hull at a fixed time $t$. Moreover, we determined the physical mechanism underlying the existence of this minimum. We showed that this latter stems from the competition between two antagonistic effects of the wall, reduction of accessible space and effective repulsion, that have separate impacts on the two complementary inward and outward parts of the convex hull. Furthermore, we considered a second subdivision of the convex hull into two complementary parts by providing an exact expression of the mean extension of the Brownian motion on the reflecting wall, which represents the length of the part of the convex hull that is along the wall.

The problem studied here can be transposed to other dimensions, e.g. the mean span of a one-dimensional Brownian motion in the presence of a reflecting point or the mean surface of the convex hull of a three-dimensional Brownian motion, that is in this case a polytope, in the presence of a reflecting plane. The two antagonistic effects of the wall that we unveiled in this paper are not limited to the dimension two and are still at play in these situations. However, it has been shown \cite{Chupeau:2015b} that in the one-dimensional version of our problem, the minimum of the mean span with respect to the rescaled distance $x$ is actually located in $x=0$, the effective repulsion being too weak to compensate the reduction of accesible space at small distance $x$. In the 2D case studied in this article, the effective repulsion is more substantial, yielding a minimum in the mean perimeter for a non-zero value of the distance. One could then expect this effect to be even more substantial in 3D, and therefore produce a minimum of the mean surface of the convex hull more marked than in 2D. Nevertheless, the proof of this conjecture requires important modifications in the calculation and in the numerical simulation, as both Cauchy formula and Graham scan algorithm would have to be adapted to the dimension three, and thus remains a open question.

Support from European Research Council starting Grant No. FPTOpt-277998 is 
acknowledged. SNM acknowledges support by ANR grant
2011-BS04-013-01 WALKMAT.

\bibliographystyle{ieeetr}

\begin{thebibliography}{10}

\bibitem{Berg:1983}
H.~C. Berg, {\em Random walks in biology}.
\newblock Princeton University Press, 1993.

\bibitem{Bartumeus:2005}
F.~Bartumeus, M.~G.~E. da~Luz, G.~Viswanathan, and J.~Catalan, ``Animal search
  strategies: a quantitative random-walk analysis,'' {\em Ecology}, vol.~86,
  no.~11, pp.~3078--3087, 2005.

\bibitem{Murphy:1992}
D.~D. Murphy and B.~R. Noon, ``Integrating scientific methods with habitat
  conservation planning: reserve design for northern spotted owls,'' {\em
  Ecological Applications}, pp.~4--17, 1992.

\bibitem{Worton:1995}
B.~J. Worton, ``A convex hull-based estimator of home-range size,'' {\em
  Biometrics}, pp.~1206--1215, 1995.

\bibitem{Giuggioli:2011}
L.~Giuggioli, J.~R. Potts, and S.~Harris, ``Animal interactions and the
  emergence of territoriality,'' {\em PLoS computational biology}, vol.~7,
  no.~3, p.~e1002008, 2011.

\bibitem{MajumdarPRL09}
J.~Randon-Furling, S.~N. Majumdar, and A.~Comtet, ``Convex hull of n planar
  brownian motions: Exact results and an application to ecology,'' {\em Phys.
  Rev. Lett.}, vol.~103, p.~140602, Sep 2009.

\bibitem{Majumdar:2010}
S.~N. Majumdar, A.~Comtet, and J.~Randon-Furling, ``Random convex hulls and
  extreme value statistics,'' {\em Journal of Statistical Physics}, vol.~138,
  no.~6, pp.~955--1009, 2010.

\bibitem{Reymbaut:2011}
A.~Reymbaut, S.~N. Majumdar, and A.~Rosso, ``The convex hull for a random
  acceleration process in two dimensions,'' {\em Journal of Physics A:
  Mathematical and Theoretical}, vol.~44, no.~41, p.~415001, 2011.

\bibitem{Dumonteil:2013}
E.~Dumonteil, S.~N. Majumdar, A.~Rosso, and A.~Zoia, ``Spatial extent of an
  outbreak in animal epidemics,'' {\em Proceedings of the National Academy of
  Sciences}, vol.~110, no.~11, pp.~4239--4244, 2013.

\bibitem{Lukovic:2013}
M.~Lukovi{\'c}, T.~Geisel, and S.~Eule, ``Area and perimeter covered by
  anomalous diffusion processes,'' {\em New Journal of Physics}, vol.~15,
  no.~6, p.~063034, 2013.

\bibitem{Randon:2013}
J.~Randon-Furling, ``Convex hull of n planar brownian paths: an exact formula
  for the average number of edges,'' {\em Journal of Physics A: Mathematical
  and Theoretical}, vol.~46, no.~1, p.~015004, 2013.

\bibitem{Randon:2014}
J.~Randon-Furling, ``Universality and time-scale invariance for the shape of
  planar l{\'e}vy processes,'' {\em Physical Review E}, vol.~89, no.~5,
  p.~052112, 2014.

\bibitem{Takacs:1980}
L.~Tak{\'a}cs, ``Expected perimeter length,'' 1980.

\bibitem{ElBachir:1983}
M.~E. Bachir, {\em L'enveloppe convexe du mouvement brownien}.
\newblock PhD thesis, Universit{\'e} Paul Sabatier, Toulouse, France, 1983.

\bibitem{Letac:1993}
G.~Letac, ``An explicit calculation of the mean of the perimeter of the convex
  hull of a plane random walk,'' {\em Journal of Theoretical Probability},
  vol.~6, no.~2, pp.~385--387, 1993.

\bibitem{Biane:2011}
P.~Biane and G.~Letac, ``The mean perimeter of some random plane convex sets
  generated by a brownian motion,'' {\em Journal of Theoretical Probability},
  vol.~24, no.~2, pp.~330--341, 2011.

\bibitem{Eldan:2014}
R.~Eldan, ``Volumetric properties of the convex hull of an n-dimensional
  brownian motion,'' {\em Electron. J. Probab}, vol.~19, no.~45, pp.~1--34,
  2014.

\bibitem{Kampf:2012}
J.~Kampf, G.~Last, and I.~Molchanov, ``On the convex hull of symmetric stable
  processes,'' {\em Proceedings of the American Mathematical Society},
  vol.~140, no.~7, pp.~2527--2535, 2012.

\bibitem{Kabluchko:2014}
Z.~Kabluchko and D.~Zaporozhets, ``Intrinsic volumes of sobolev balls,'' {\em
  arXiv preprint arXiv:1404.6113}, 2014.

\bibitem{Chupeau:2015b}
M.~Chupeau, O.~B\'enichou, and S.~N. Majumdar, ``Convex hull of a brownian
  motion in confinement,'' {\em Phys. Rev. E}, vol.~91, p.~050104, 2015.

\bibitem{Prudnikov3}
Y.~A. Brychkov and A.~Prudnikov, {\em Integral transforms of generalized
  functions}.
\newblock Taylor and Francis, 1986.

\bibitem{Abramowitz}
M.~Abramowitz and I.~A. Stegun, {\em Handbook of mathematical functions: with
  formulas, graphs, and mathematical tables}.
\newblock Courier Dover Publications, 2012.

\bibitem{Prudnikov2}
A.~P. Prudnikov, Y.~A. Brychkov, and O.~I. Marichev, {\em Integrals and
  Series.}, vol.~2. Special Functions.
\newblock Taylor and Francis, 1983.

\bibitem{Chupeau:2015a}
M.~Chupeau, O.~B{\'e}nichou, and S.~N. Majumdar, ``Survival probability of a
  brownian motion in a planar wedge of arbitrary angle,'' {\em Physical Review
  E}, vol.~91, no.~3, 2015.

\bibitem{Graham:1972}
R.~L. Graham, ``An efficient algorith for determining the convex hull of a
  finite planar set,'' {\em Information processing letters}, vol.~1, no.~4,
  pp.~132--133, 1972.

\bibitem{randon:these}
J.~Randon-Furling, {\em Statistiques d'extr{\^e}mes du mouvement brownien et
  applications}.
\newblock PhD thesis, Universit{\'e} Paris Sud-Paris XI, 2009.

\bibitem{Redner}
S.~Redner, {\em A guide to first-passage processes}.
\newblock Cambridge University Press, 2001.

\bibitem{Bray:2013}
A.~J. Bray, S.~N. Majumdar, and G.~Schehr, ``Persistence and first-passage
  properties in nonequilibrium systems,'' {\em Advances in Physics}, vol.~62,
  no.~3, pp.~225--361, 2013.

\end{thebibliography}

\appendix

\section{Survival probability in a planar infinite absorbing wedge}\label{Survival}

\begin{figure}[H]
\centering
\includegraphics[width=120pt]{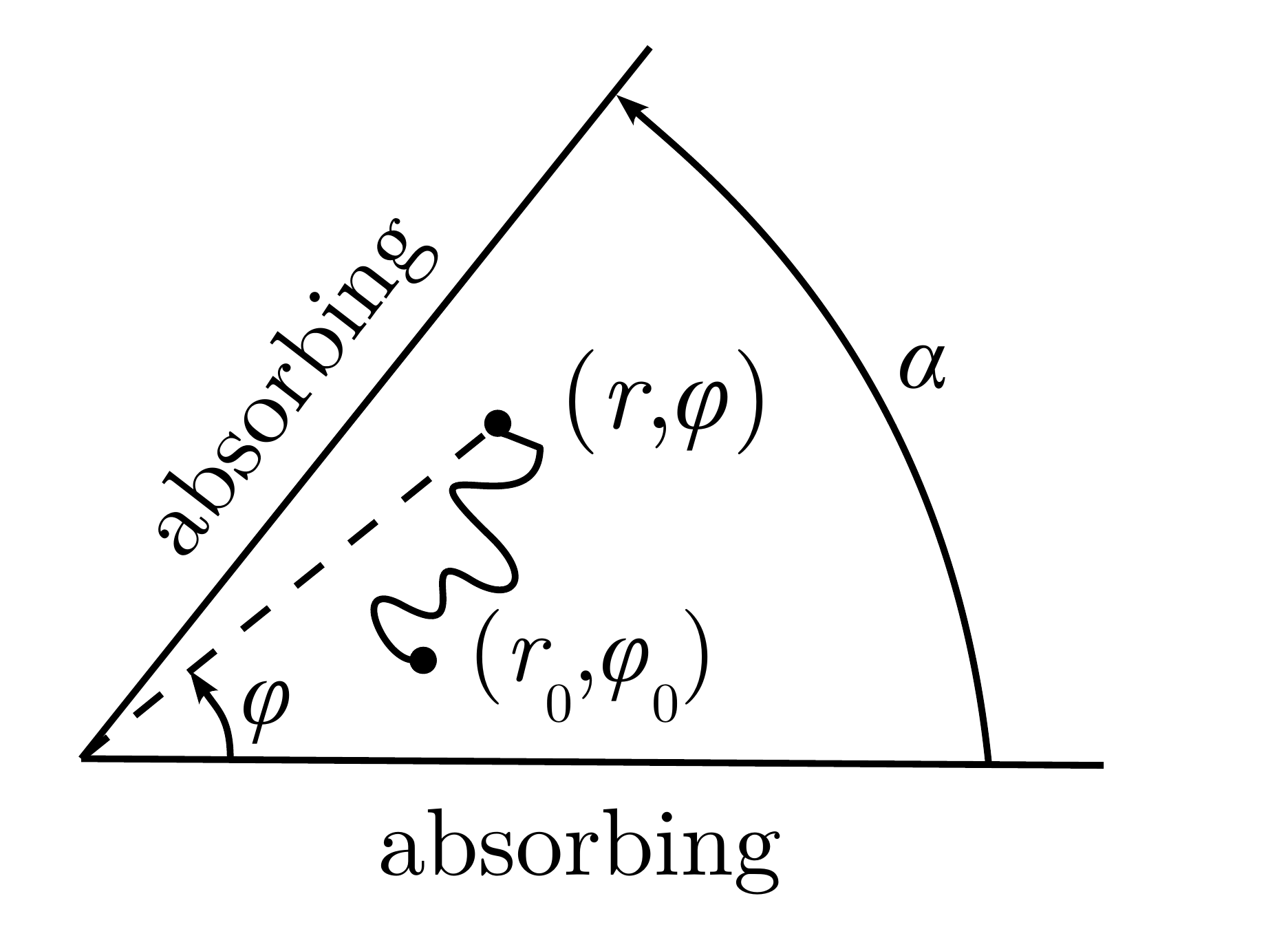}
\caption{Brownian motion in an infinite absorbing planar wedge of top angle $\alpha$, described in polar coordinates and starting from the point \mbox{$(r_0,\varphi_0)$}.}
\label{coord}
\end{figure}

We determine the survival probability of a Brownian walker in a planar infinite absorbing wedge. We first solve the backward Fokker-Planck equation for the propagator \mbox{$P(r,\varphi,t|r_0,\varphi_0,0)$} in polar coordinates (see Fig.~\ref{coord}), where the origin is set at the apex of the wedge. The Brownian motion starts at \mbox{$(r_0,\varphi_0)$} at time $0$, so when there is no ambiguity, we refer to the propagator as \mbox{$P(r,\varphi,t)$}
\begin{equation}
\frac{\partial P}{\partial t}=D \Delta P= D \left[ \frac{\partial^2 P}{\partial r^2} + \frac{1}{r}\frac{\partial P}{\partial r}+ \frac{1}{r^2} \frac{\partial^2 P}{\partial \varphi^2} \right]
\end{equation}
with the following initial and boundary conditions
\begin{align} \label{backward}
&P(r,\varphi,0)=\delta(\bm{r}-\bm{r_0})=\frac{1}{r_0} \delta(r-r_0) \delta(\varphi-\varphi_0) \\
&P(r,0,t)=0\\
&P(r,\alpha,t)=0.
\end{align}
The Laplace transform $\hat{P}$ of the propagator is defined by
\begin{equation}
\hat{P}(r,\varphi,s)=\int_0^{+\infty} dt \; P(r,\varphi,t) \; e^{-st}.
\end{equation}
If we apply Laplace transform to Eq.~(\ref{backward}), we get
\begin{equation} \label{eqTL}
\displaystyle\int_0^{+\infty} dt \; \frac{\partial P}{\partial t} \; e^{-st}  =D \Delta \hat{P}
\end{equation}
Integrating the left-hand side by parts, we obtain
\begin{equation}
s \hat{P} - \frac{1}{r_0} \delta(r-r_0) \delta(\varphi-\varphi_0)=D \Delta \hat{P}
\end{equation}
with the boundary conditions
\begin{equation}\label{CL}
\hat{P}(r,0,s)=\hat{P}(r,\alpha,s)=0.
\end{equation}

Eq.~\eqref{CL} admits solutions of the form \mbox{$R(r,s) \Phi(\varphi,s)$}. Plugging this expression into \eqref{eqTL} and separating $r$-dependent and $\varphi$-dependent terms, we get
\begin{equation}
 r^2 \left( \frac{s}{D}- \frac{R''}{R} \right)- r \frac{R'}{R} = \frac{\Phi''}{\Phi}.
\end{equation}
The angular boundary conditions \eqref{CL} indicate that \mbox{$\Phi(\varphi,s)$} is a linear combination of \mbox{$\sin(n\pi \varphi/\alpha)$}. As Dirac delta function can be written
\begin{equation}
\delta(\varphi-\varphi_0)=\frac{2}{\alpha} \sum\limits_{n=1}^{+\infty} \sin \left(n\frac{\pi \varphi}{\alpha}\right) \sin\left(n\frac{\pi \varphi_0}{\alpha}\right),
\end{equation}
we can finally decompose $\hat{P}$ on the same basis
\begin{equation}
\hat{P}(r,\varphi,s)=\sum\limits_{n=1}^{+\infty} R_n(r,s) \sin \left(n\frac{\pi \varphi}{\alpha}\right) \sin\left(n\frac{\pi \varphi_0}{\alpha}\right).
\end{equation}
Let us plug this expression into \eqref{eqTL}. Each component \mbox{$R_n(r,s)$} satisfies
\begin{equation}
s R_n-\frac{2}{r_0 \alpha} \delta(r-r_0)=D \left(R''_n+\frac{1}{r} R'_n-\frac{k_n^2}{r^2}  R_n \right).
\end{equation}
 with \mbox{$k_n=n\pi/\alpha$}. Introducing the variable \mbox{$y=r \sqrt{s/D} $}, we obtain the following modified Bessel equation
\begin{equation}\label{EqRn}
R''_n(y)+\frac{1}{y} R'_n(y) -\left(1+\frac{k_n^2}{y^2} \right) R_n(y)=-\frac{2}{y_0 D \alpha} \delta(y-y_0).
\end{equation}
$R_n(y,s)$ remains finite, so we write
\begin{equation}
\begin{cases}
R_n(y,s)=A_n I_{k_n}(y)  \qquad \text{for}  \quad y<y_0\\
R_n(y,s)=B_n K_{k_n}(y)  \qquad \text{for}  \quad y>y_0
\end{cases}
\end{equation}
with \mbox{$I_k(y)$} and \mbox{$K_k(y)$} modified Bessel functions of first and second order. We determine the two constants $A_n$ and $B_n$ by writing, on the one hand, the continuity of \mbox{$R_n(y,s)$} at $y_0$, and on the other hand by integrating \eqref{EqRn} between \mbox{$y_0^-$} and \mbox{$y_0^+$}. This yields
\begin{equation}
\begin{dcases}
A_n I_{k_n}(y_0)=B_n K_{k_n}(y_0)\\
B_n K'_{k_n}(y_0)-A_n I'_{k_n}(y_0)= -\frac{2}{\alpha D y_0}.
\end{dcases}
\end{equation}
Writing the Wronskian 
\begin{equation}
W(K_k,I_k)=K_k(y)I'_k(y)-K'_k(y)I_k(y)=\frac{1}{y},
\end{equation}
we obtain
\begin{equation}
\begin{dcases}
A_n=\frac{2}{\alpha D} K_{k_n}(y_0)\\[.9em]
B_n=\frac{2}{\alpha D} I_{k_n}(y_0).
\end{dcases}
\end{equation}
This yields
\begin{align}
&\hat{P}(r,\varphi,s)=\frac{2}{\alpha D} \sum\limits_{n=1}^{+\infty} I_{\frac{n\pi}{\alpha}} \left( \sqrt{\frac{s}{D}} \min(r_0,r)\right)  \nonumber \\
& \times K_{\frac{n\pi}{\alpha}}  \left( \sqrt{\frac{s}{D}} \max(r_0,r) \right) \sin\left(\frac{n \pi \varphi}{\alpha}\right) \sin\left(\frac{n \pi \varphi_0}{\alpha}\right) \nonumber\\
\end{align}
and by taking the inverse Laplace transform, we eventually obtain the propagator in the wedge
\begin{align}
P(r,\varphi,t|r_0,\varphi_0)&=\frac{1}{\alpha D t} \sum\limits_{n=1}^{+\infty} \sin\left(\frac{n \pi \varphi}{\alpha} \right) \sin\left(\frac{n \pi \varphi_0}{\alpha} \right) \nonumber \\ 
&  \times I_{\frac{n\pi}{\alpha}}\left(\frac{r_0 r}{2Dt} \right) \exp \left(- \frac{r^2+r_0^2}{4Dt} \right).
\end{align}

The survival probability is then calculated from the propagator through
\begin{align}
&S(t|r_0,\varphi_0) \equiv \int_0^{+\infty} dr \; r   \int_0^{\alpha} d\varphi \; P(t,r,\varphi|r_0,\varphi_0) \nonumber \\
&\quad =\frac{1}{\alpha Dt} \sum\limits_{m=0}^{+\infty} \frac{2 \alpha}{(2m+1)\pi} \sin\left( \frac{(2m+1)\pi\varphi_0}{\alpha}\right)  \nonumber \\
& \qquad \times \int_0^{+\infty} dr \; r  \: I_{\frac{(2m+1)\pi}{\alpha}} \left(\frac{r_0 r}{2Dt} \right) e^{-\frac{r_0^2+r^2}{4Dt}}. 
\end{align}
We then integrate by parts 
\begin{align}
S(t|r_0,\varphi_0)&=\frac{2 r_0}{\alpha Dt} \sum\limits_{m=0}^{+\infty} \sin\left( \frac{(2m+1)\pi\varphi_0}{\alpha}\right) \nonumber \\
& \times \int_0^{+\infty} dr I'_{\frac{(2m+1)\pi}{\alpha}} \left(\frac{r_0 r}{2Dt} \right) e^{-\frac{r_0^2+r^2}{4Dt}}. 
\end{align}
Using the property \cite{Abramowitz} 
\begin{equation} 
I'_k(x)=\frac{I_{k-1}(x)+I_{k+1}(x)}{2}
\end{equation}
and the integral \cite{Prudnikov2}
\begin{equation}
\int_0^{+\infty} dr I_{k-1}\left( \frac{r_0 r}{2Dt} \right) e^{-\frac{r^2}{4Dt}}=\sqrt{\pi Dt} \e{\frac{r_0^2}{8 Dt}} I_{\frac{k-1}{2}}\left(\frac{r_0^2}{8Dt}\right),
\end{equation}
we obtain an expression of the survival probability at time $t$, having started at time $0$ from \mbox{$(r_0,\varphi_0)$}
\begin{align}
&S(t|r_0,\varphi_0)=\frac{r_0}{\sqrt{\pi Dt}} \;e^{-\frac{r_0^2}{8Dt}} \sum\limits_{m=0}^{+\infty} \frac{\sin\left(\frac{(2m+1)\pi \varphi_0}{\alpha}\right)}{2m+1} \nonumber \\
&\times \left[ I_{\frac{(2m+1)\pi}{2 \alpha}-\frac{1}{2}} \left(\frac{r_0^2}{8Dt}\right)+ I_{\frac{(2m+1)\pi}{2 \alpha}+\frac{1}{2}} \left(\frac{r_0^2}{8Dt}\right) \right]. 
\end{align}

\section{Geometric relations between \mbox{$(r_0,\varphi_0)$} and \mbox{$(M,\theta)$}}\label{Geometry}

\begin{figure}[h]
\centering
\includegraphics[width=120pt]{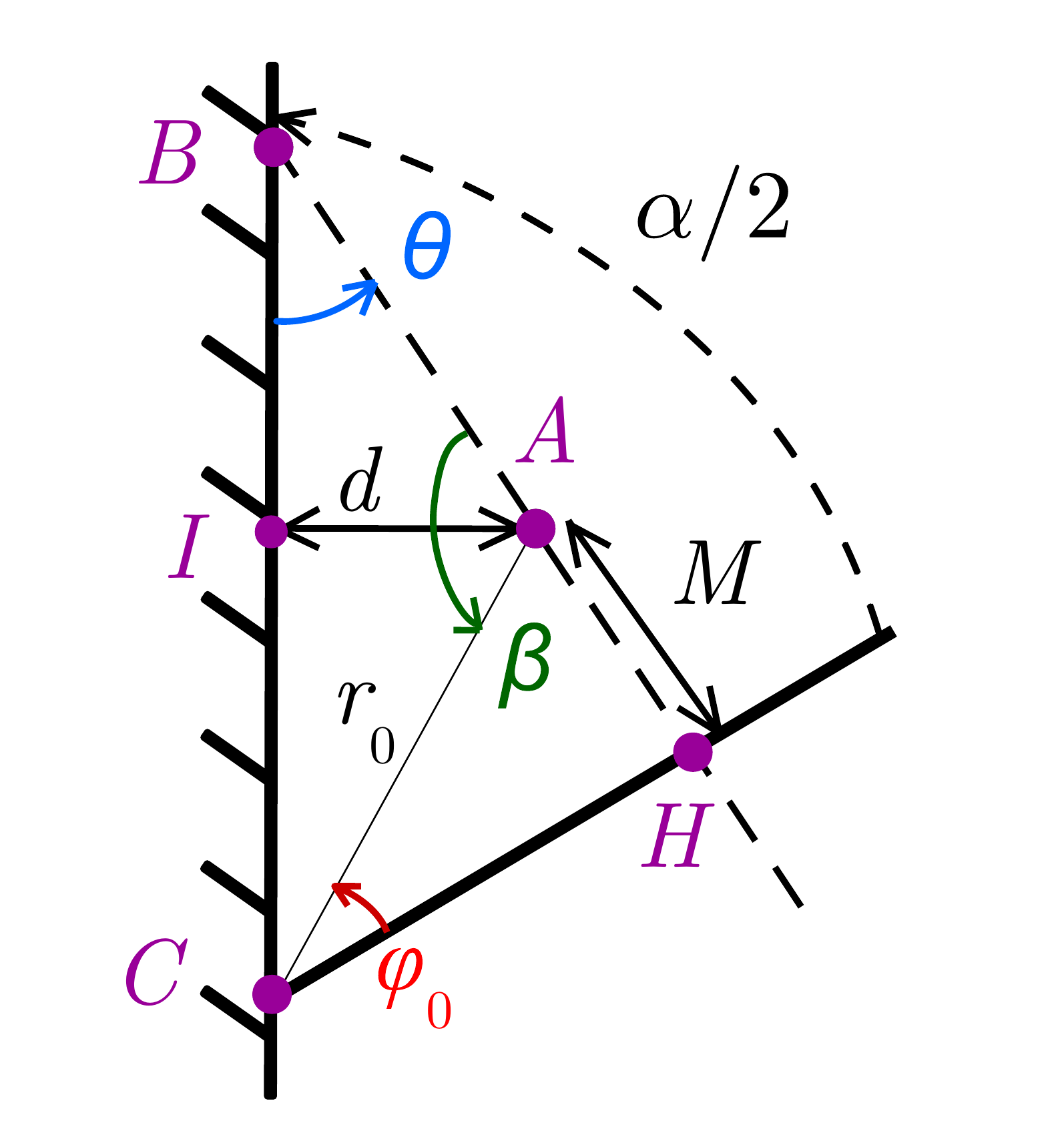}
\caption{Definition of the points, distances and angles used in the derivation of the geometric relations between \mbox{$(r_0,\varphi_0)$} and \mbox{$(M,\theta)$}.}
\label{geom_schema}
\end{figure}

Natural variables in the wedge are the polar coordinates $r_0$ and $\varphi_0$, but Eq. \eqref{defperim} requires to work with the variables $M$ and $\theta$. We then have to establish the correspondence between the two sets of coordinates \mbox{$(r_0,\varphi_0)$} and \mbox{$(M,\theta)$}.
First, we write the sum of angles in triangles $BAC$ and $AHC$ (see Fig.~\ref{geom_schema})
\begin{eqnarray}
&\theta+\beta+\frac{\alpha}{2}-\varphi_0=\pi\\
&\pi-\beta+\frac{\pi}{2}+\varphi_0=\pi.
\end{eqnarray}
Combining these two equations, we get the top angle of the wedge
\begin{equation}\label{defalpha}
\alpha=\pi-2 \theta.
\end{equation}
Then, we need to write $r_0$ and $\varphi_0$ in terms of $M$ and $\theta$. From triangle $AHC$, we get
\begin{equation}\label{defM}
M=r_0 \sin\varphi_0,
\end{equation}
and from triangle $ACI$, 
\begin{equation}\label{defd}
d=r_0 \sin\left( \frac{\alpha}{2} -\varphi_0\right)=r_0 \cos(\theta+\varphi_0).
\end{equation}
It gives
\begin{equation}
\frac{M}{\sin\varphi_0}=\frac{d}{\cos(\theta+\varphi_0)}
\end{equation}
from which we extract 
\begin{equation}\label{tantheta}
\tan\varphi_0=\frac{M \cos\theta}{d+M\sin\theta}.
\end{equation}
Then, using  \eqref{defM},
\begin{align}\label{defr0}
r_0&=\frac{M}{\sin\varphi_0}=\frac{M}{\sqrt{1-\cos^2\varphi_0}} \nonumber \\
&=\frac{M \sqrt{1+\tan^2\varphi_0}}{\tan\varphi_0}
\end{align}
and then
\begin{equation}\label{r0}
r_0=\frac{1}{\cos\theta} \sqrt{d^2+2 d M \sin\theta+M^2}.
\end{equation}
Expression (\ref{tantheta}) is not suited to obtain $\varphi_0$ because $\arctan$ gives values in \mbox{$\left[-\pi/2,\pi/2\right]$} whereas $\varphi_0$ is in the range \mbox{$\left[0,\pi\right]$}. Hence $\varphi_0$ must be determined via an $\arccos$, which gives values in the good range. Using Eqs.~(\ref{defd}) and (\ref{defM})
\begin{equation}
\frac{d}{r_0}=\cos(\theta+\varphi_0)=\cos\theta\cos\varphi_0-\sin\theta \; \frac{M}{r_0}
\end{equation}
and getting rid of $r_0$ with  \eqref{defr0}, we obtain the expected relation
\begin{equation}\label{phi0}
\varphi_0=\arccos\left( \frac{d+M \sin\theta}{\sqrt{M^2+2dM\sin\theta+d^2}}\right).
\end{equation}
These relations remain true if $\theta$ is negative.

\end{document}